\documentclass[onecolumn,superscriptaddress,amsmath,amssymb,prb]{revtex4}
\usepackage{psfrag,graphicx,stmaryrd,bm}
\newcommand {\bb}{\bibitem}
\newcommand {\be}{\begin{equation}}
\newcommand {\ee}{\end{equation}}
\newcommand {\bea}{\begin{eqnarray}}
\newcommand {\eea}{\end{eqnarray}}

\begin{document}


\title{BCS theory of nodal superconductors}

\author{Hyekyung Won}
 \affiliation{Department of Physics, Hallym University, Chuncheon, 200-702, South Korea}
\author{Stephan Haas}
 \affiliation{Department of Physics and Astronomy, University of Southern California, Los Angeles,
 California 90089-0484, USA}
\author{David Parker}
 \affiliation{Department of Physics and Astronomy, University of Southern California, Los Angeles,
 California 90089-0484, USA}
\author{Sachin Telang}
 \affiliation{Department of Physics and Astronomy, University of Southern California, Los Angeles,
 California 90089-0484, USA}
\author{Andr\'as V\'anyolos}
 \affiliation{Department of Physics, Budapest University of Technology and Economics, H-1521 Budapest, Hungary}
\author{Kazumi Maki}
 \affiliation{Department of Physics and Astronomy, University of Southern California, Los Angeles,
 California 90089-0484, USA}

\date{\today}

\begin{abstract}
This course has a dual purpose. First we review the successes of the weak-coupling BCS theory 
in describing new classes of superconductors discovered since 1979. They include the heavy-fermion 
superconductors, organic superconductors, high-$T_c$ cuprate superconductors, Sr$_2$RuO$_4$ etc.
Second, we present the quasiclassical approximation introduced by Volovik, which we extend to describe
the thermodynamics and the thermal conductivity of the vortex state in nodal superconductors.  This approach provides
the most powerful tool in identifying the symmetry of the energy gap function $\Delta(\mathbf{k})$
in these new superconductors.
\end{abstract}

\maketitle

\section{Introduction}
\noindent
Question: \emph{What is the difference between a Fermi liquid and a non-Fermi liquid?}

\noindent
Answer: \emph{The difference is the same as the one between bananas and non-bananas.}
\vskip 0.5cm
\hfill Boris Altschuler (2001)

\vskip 0.5cm
Unconventional or ``nodal'' superconductors appeared on the scene in 1979, when the heavy-fermion
superconductor CeCu$_2$Si$_2$ and the organic superconducting Bechgaard salts (TMTSF)$_2$PF$_6$ were
discovered. Since then, many more heavy-fermion superconductors [\onlinecite{sigrist}] and organic superconductors
[\onlinecite{ishizuro}] have been
synthesized. This development was followed by the epoch making discovery of high-$T_c$ cuprate superconductor
La$_{2-x}$Ba$_x$CuO$_4$ by Bednorz and
M\"uller [\onlinecite{bednorz}] with the superconducting transition temperature $T_c=35$K in 1986. Within 
a few years new classes of high-$T_c$ cuprates emerged, including YBa$_2$Cu$_3$O$_{6+\delta}$,
La$_{2-x}$Sr$_x$CuO$_4$, Bi$_{2}$Sr$_{2}$Ca$_{1-x}$Y$_{x}$Cu$_{2}$O$_{8+\delta}$ (Bi2212);
and HgBa$_{2}$CaCu$_{2}$O$_{6}$ with $T_c=145$K. The subsequent enthusiasm
and confusion are well documented in an early review by Enz [\onlinecite{charles}]. Confusion? 
Yes, initially it was thought that Landau's Fermi liquid theory [\onlinecite{landau}] and the BCS theory
[\onlinecite{bardeen}] were no longer applicable.

Among many proposals one of the most influential were Anderson's dogmas [\onlinecite{anderson-1}], which can be summarized
as follows:
\begin{enumerate}
\item[a.] The action takes place in the CuO$_2$ plane common to all high-$T_c$ cuprate superconductors.

\item[b.] The undoped state is a Mott insulator with antiferromagnetic (AF) order. Upon
doping superconductivity appears. Therefore the simplest
Hamiltonian is the two-dimensional (2D) one-band Hubbard model:

\begin{equation}
H=-t\sum_{\langle i,j\rangle\alpha}(c_{i\alpha}^+c_{j\alpha}+ h.c)+U\sum_{i}n_{i\shortuparrow}n_{i\shortdownarrow},
\label{hamiltonian}
\end{equation}
where $\langle i,j\rangle$ connects the nearest neighbors in the 2D square lattice. The $c_{i\alpha}^+$ and
$c_{i\alpha}$ are creation and annihilation operator for the holes at the site $i$ with spin $\alpha$
and $n_{i\shortuparrow}=c_{i\shortuparrow}^+c_{i\shortuparrow}$.

\item[c.] As a possible ground state of Eq.~\eqref{hamiltonian} Anderson proposed the resonating valence bond
(RVB) state:

\begin{equation}
\Psi=\prod_{i}(1-d_i)|\text{BCS}\rangle,\label{gutzwiller}
\end{equation}
where $|\text{BCS}\rangle$ is the BCS state for $s$-wave superconductors [\onlinecite{bardeen}], and $\prod_i(1-d_i)$
with $d_i=n_{i\shortuparrow}n_{i\shortdownarrow}$ is called the Gutzwiller operator. $\prod_i(1-d_i)$ 
annihilates all doubly occupied states.
\end{enumerate}

In spite of tremendous efforts spent on both Eqs .~\eqref{hamiltonian} and ~\eqref{gutzwiller} it has been
difficult to find solutions in two dimensions.  
On the other hand the 1D version of Eq.~\eqref{hamiltonian} is now completely understood
[\onlinecite{lieb,ogata-1}].  In the meantime the perturbative analyses 
based on Eq.~\eqref{hamiltonian} predict BCS $d$-wave
superconductivity in high-$T_c$ cuprates [\onlinecite{moriya,monthoux,scalapino,pao}].

High-quality single crystals of YBCO, LSCO and thin films of Bi2212 became available around 1992. The
$d$-wave superconductivity in these high-$T_c$ cuprates was established in 1994. Among many
experiments, angle-resolved photoemission spectroscopy (ARPES) [\onlinecite{damascelli}] and
Josephson interferometry [\onlinecite{harlingen,tsuei}] played a crucial role.  Around this time
several theoretical groups started to
analyze the physical properties of $d$-wave superconductivity within the BCS framework
[\onlinecite{won-1,sun-1,sun-2,maki-1}]. In 1993 Patrick Lee [\onlinecite{lee-1}] demonstrated the universal
heat conduction in $d$-wave superconductors. Furthermore the thermal conductivity was shown to increase with
increasing impurity scattering [\onlinecite{sun-2}]. This counterintuitive behavior was observed in the 
Zn-substituted YBCO in [\onlinecite{taillefer}]. The electronic contribution to the thermal conductivity 
in $d$-wave superconductors
is proportional to $T$ at low temperatures (i.e. $T\ll\Delta$ where $\Delta$ is the maximum value
of the energy gap). Here we assume $\Delta(\mathbf{k})=\Delta\cos(2\phi)$ and $\phi$ is the angle $\mathbf{k}$
in the $a-b$ plane makes from the $a$ axis. Then in the limit of no impurity scattering (i.e. $\Gamma\to0$ where $\Gamma$
is the quasiparticle scattering rate in the normal state) the thermal conductivity is given by

\begin{equation}
\frac{\kappa_{00}}{T}=\frac{k_B^2}{3\hbar}\frac{v}{v_2}n,\label{thermal}
\end{equation}
where $\kappa_{00}/T=\lim_{\Gamma\to0}\kappa/T$, and $v/v_2=E_F/\Delta$ and $n$ is the quasiparticle
density or the hole density. The velocities $v$ and $v_2$ are defined 
from the quasiparticle energy at the Dirac cone

\begin{equation}
E_{\mathbf{k}}=\sqrt{v^2(k_\parallel-k_F)^2+v_2^2k_\perp^2},\label{dirac}
\end{equation}
where $v$ is the Fermi velocity, $k_\parallel$ is the radial component of the wave vector and $k_\perp$
the component perpendicular to the Fermi surface.

Later we shall derive Eq.(3) in Section IV. The thermal conductivity in single
crystals of optimally doped YBCO and Bi2212 below $T=1$K was measured by May Chiao et al
[\onlinecite{chiao-1,chiao-2}].
Making use of Eq.~\eqref{thermal} they found $\Delta/E_F=1/10$ and $1/14$ for Bi2212 and YBCO respectively.
These remarkable ratios imply:
\begin{enumerate}
\item[a.] High-$T_c$ cuprates are described by the BCS theory of $d$-wave
superconductivity.  They are far away from the Bose-Einstein condensation limit which requires $\Delta\sim E_F$
[\onlinecite{micnas}].

\item[b.] According to the Ginzburg criterion, fluctuation effects are of the order 
$\Delta/E_F\sim(p_F\xi)^{-1}$. In other words they should be at most 10\%.
This appears to exclude large phase fluctuations and stripe phases discussed in
[\onlinecite{orenstein,kivelson}].

\item[c.] For $\Delta/E_F=1/10$ there are hundreds of quasiparticle bound states around the core of
a single vortex in $d$-wave superconductors [\onlinecite{kato-1,kato-2}]. The radial ($\mathbf{r}$)
dependence of the local quasiparticle density of states is very similar to the one obtained for $s$-wave
superconductivity [\onlinecite{gygi}]. We show in Fig. 1 the local density of states
around a single vortex for a $d$-wave and $s$-wave superconductor.  These are well-known bound
states first discovered by Caroli, de-Gennes and 
Matricon [\onlinecite{caroli-1,caroli-2}] for $s$-wave superconductors.
\end{enumerate}

\begin{figure}[t!]
\includegraphics[width=7cm]{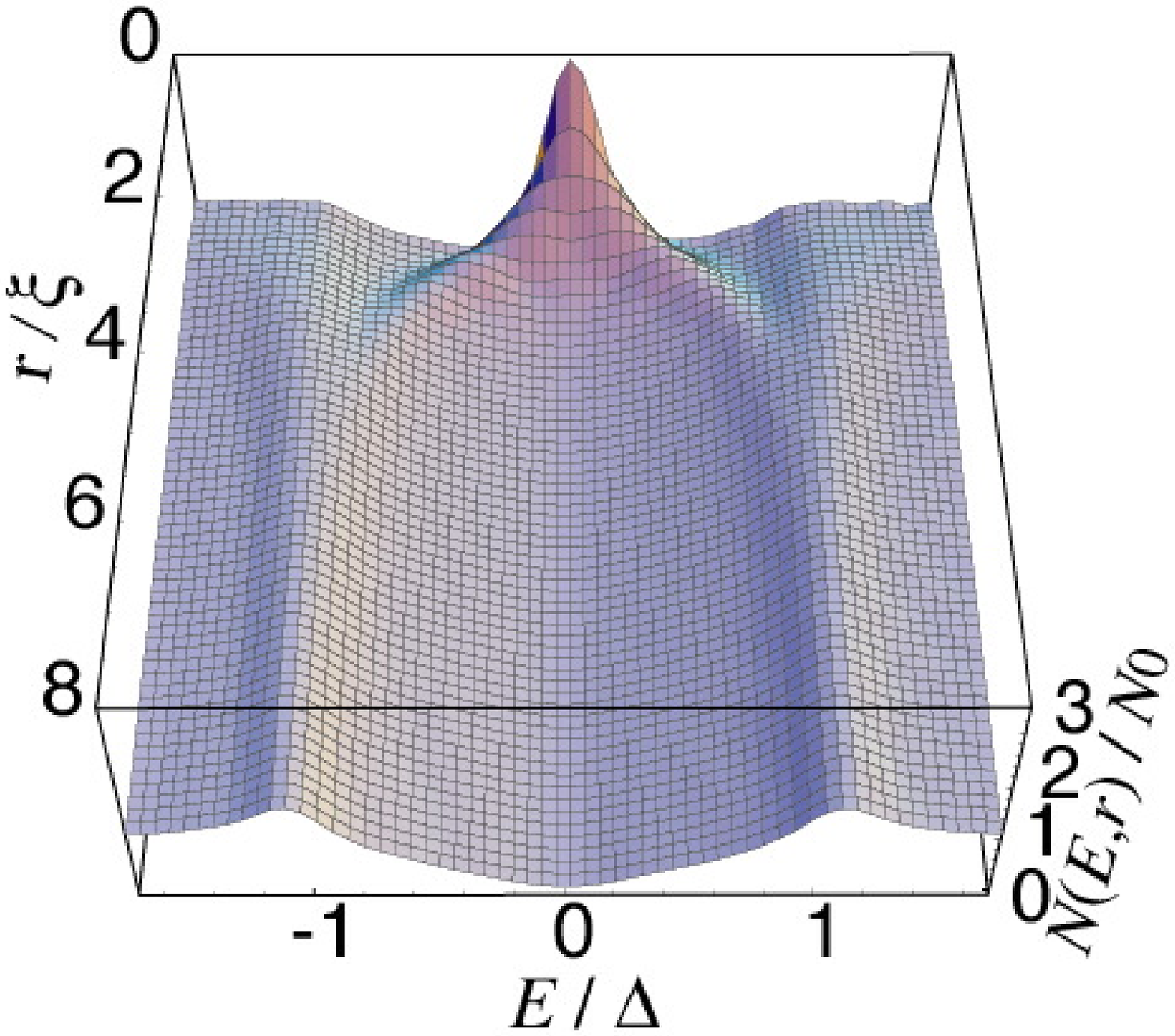}\hspace{1cm}
\includegraphics[width=7cm]{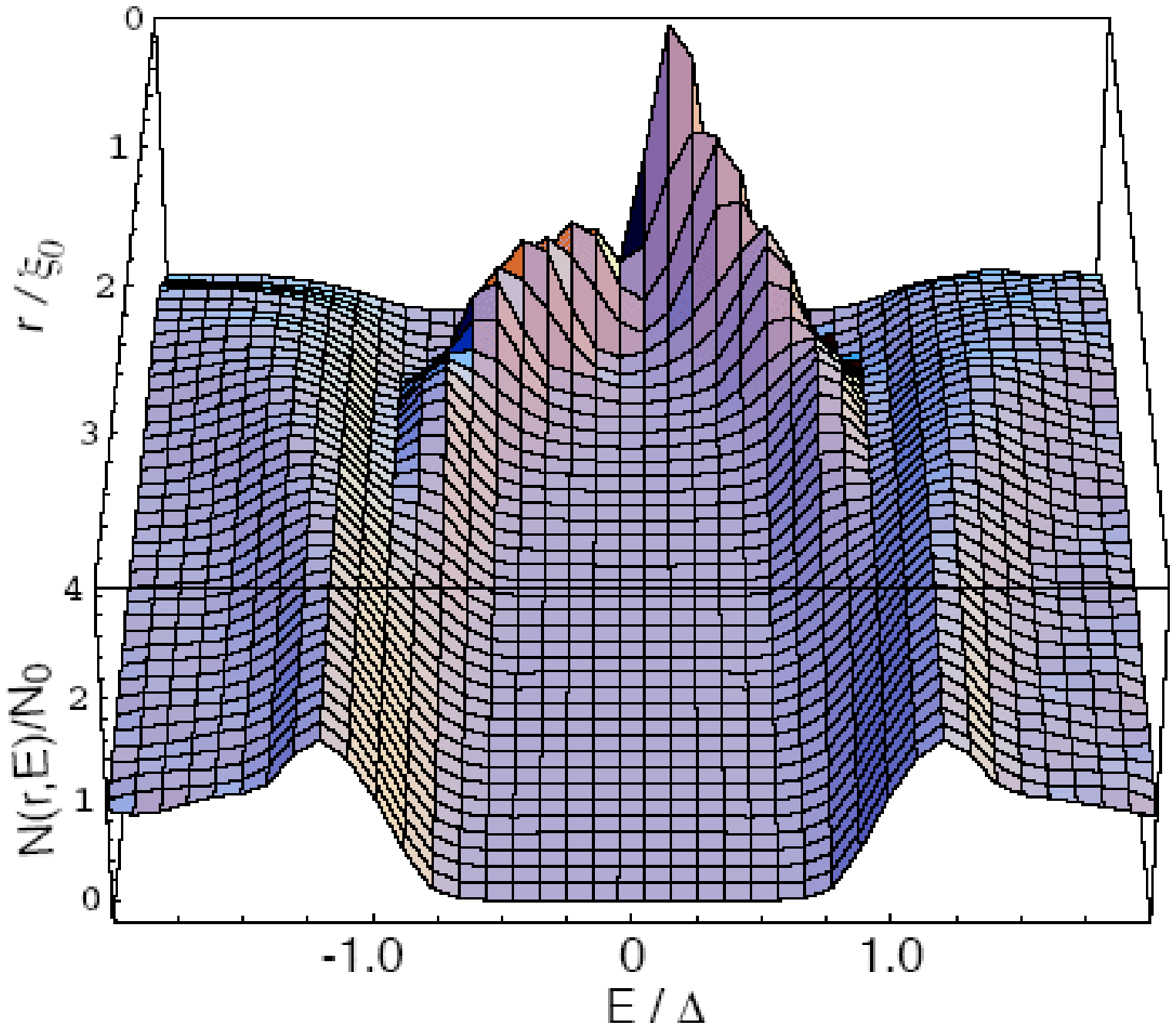}
\caption{\label{local-dos} Local density of states around a single vortex for d-wave (left) and
s-wave superconductivity.}
\end{figure}

\noindent
In earlier works [\onlinecite{franz,takizawa,yasni,ogata-2}] it was asserted that
there would be no bound state around a single vortex in $d$-wave superconductors. However in these works
it was assumed $\Delta\simeq E_F$ in order to facilitate the numerical analysis based on the lattice version
of the Bogoliubov de Gennes equation.  This
assumption ($\Delta\simeq E_F$) eliminates the bound states in these numerical analyses.
If we limit ourselves to the experiments on single crystals of high-T$_{c}$
cuprate superconductors, we find that most of the behaviors observed are consistent with the
BCS theory of d-wave superconductivity [\onlinecite{hussey}].  Also it is better to use
the continuum version of the  Bogoliubov de Gennes equation.  which is 
proposed in [\onlinecite{morita}] and used in [\onlinecite{kato-1,kato-2}].  More recently a similar
analysis is extended for a vortex in an f-wave superconductor [\onlinecite{kato-new}].  

From a theoretical point of view the universality of the Landau Fermi liquid in 2D systems was
demonstrated within the renormalization group analysis [\onlinecite{shankar,metzner,houghton}].
The quasiparticles in the normal state of high-$T_c$ cuprates appear to be a Fermi liquid state.
Furthermore the quasiparticles in $d$-wave superconductors are in a BCS-Fermi liquid state with the quasiparticle energy

\begin{equation}
E_{\mathbf{k}}=\sqrt{v^2(k_\parallel-k_F)^2+\Delta^2\cos^2(2\phi)}.\label{quasi}
\end{equation}
In the vicinity of the Dirac cone Eq.~\eqref{quasi} reduces to Eq.~\eqref{dirac}.

\begin{figure}[t!]
\psfrag{x}{$x$}
\psfrag{T}{$T$}
\psfrag{t2}{$T_c$}
\psfrag{t1}{$T^*$}
\psfrag{af}{AF}
\psfrag{d-wave}{$d$-wave SC}
\psfrag{pseudo}{pseudogap phase}
\includegraphics[width=8cm]{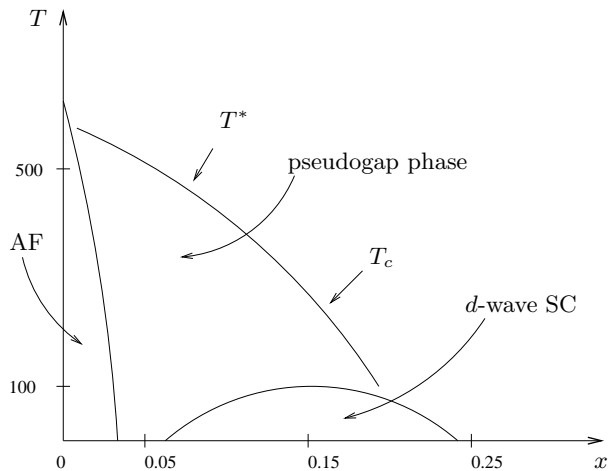}
\caption{\label{phase-diagram} The phase diagram of hole-doped high-T$_{c}$ cuprate superconductors.}
\end{figure}

Another consequence of the renormalization group analysis is that the instability of the normal Fermi liquid
is related to the infrared divergence in the particle-particle (and the hole-hole) channel or the particle-hole
channel. The former results in conventional or unconventional superconductivity and the latter in conventional
or unconventional density wave states. Therefore it is of great interest to look at the phase diagram 
of the high-$T_c$ cuprates
from this point of view. We show a schematic phase diagram in Fig.~\ref{phase-diagram}.

Let us consider the hole-doped region.
In the vicinity of $x=0$ there is an antiferromagnetic (AF) insulating phase, which is a Mott
insulator (MI). As the hole-doping $x$ is increased, the AF order is rapidly suppressed around $x=3\%$. Then a
superconducting region develops for $5\% < x < 25\%$.  This is sometimes called ``the superconducting dome".

Also in the underdoped region there is the pseudogap (PG) regime. 
The nature of the pseudogap is still hotly debated.  There is evidence that it is a $d$-wave density wave (dDW)
[\onlinecite{cappelutti,benfatto,chakravarty,dora-1}] with an energy gap $\Delta(\mathbf{k})=\Delta\cos(2\phi)$.
Earlier proposals [\onlinecite{cappelutti,chakravarty}] have only considered the commensurate case with $Z_2$
symmetry something similar to a flux phase [\onlinecite{affleck,senthil}].

On the contrary, we consider the incommensurate $d$-wave density wave [\onlinecite{benfatto,dora-1}] case
where the condensate has U(1) symmetry as in conventional charge-density wave systems (CDW).  Here the phase vortex [\onlinecite{ong}]
is the most common topological defect.  Moreover the phase diagram suggests a coexistence region of dDW 
and $d$-wave superconductivity
[\onlinecite{cappelutti,chakravarty}]. Then the phenomenological gap introduced by Tallon and Loram
[\onlinecite{tallon}] should be the energy gap associated with dDW.

Recently Laughlin [\onlinecite{laughlin}] has pointed out that the wave function (2) is impractical, since
the Gutzwiller operator has no inverse. Instead, he proposed to analyze Eq.~\eqref{gutzwiller} with a modified Gutzwiller
operator which has an inverse. For example

\begin{equation}
\prod_i(1-d_i) \quad \to \quad \prod_i(1-\alpha d_i),\label{modified-gutzwiller}
\end{equation}
with $\alpha<1$. He called this ``gossamer superconductivity". ``Gossamer" means filmy cobweb
or something light, fragile but strong.  If we look at this new wave function with Eq.~\eqref{modified-gutzwiller}, 
we realize that this is an example of
competing order parameters [\onlinecite{haas,won-2}]. We shall come back to this question at the end of our course.

\begin{figure}[t!]
\includegraphics[width=4cm,height=5.5cm]{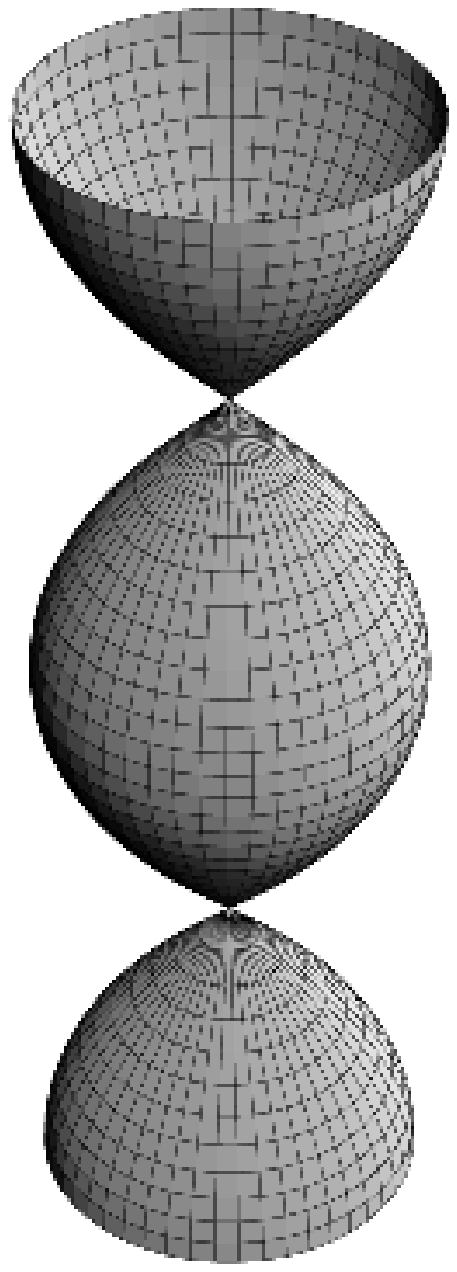}
\includegraphics[width=6cm]{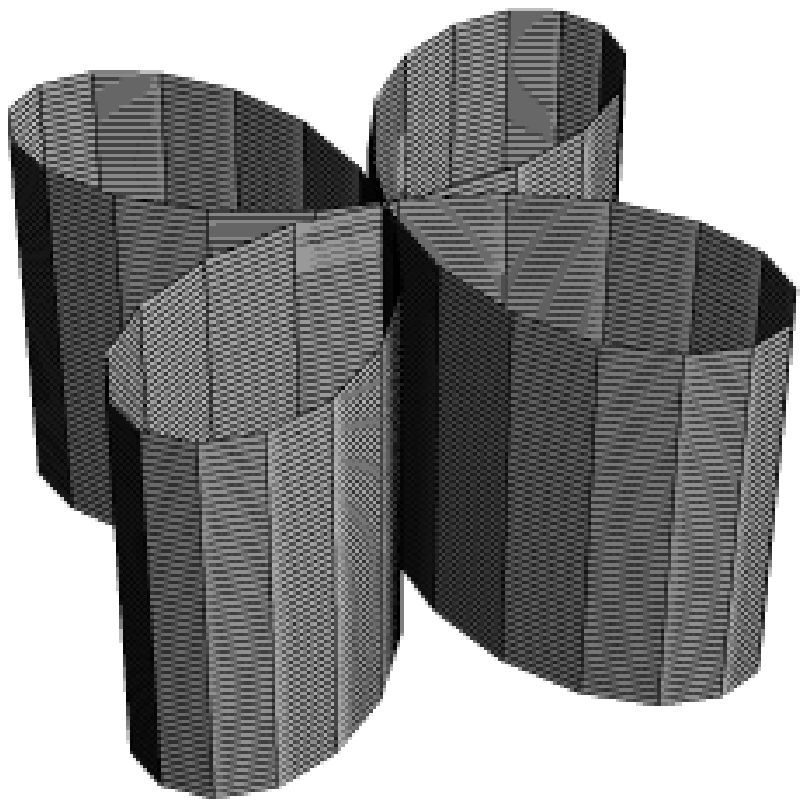}
\includegraphics[width=7cm]{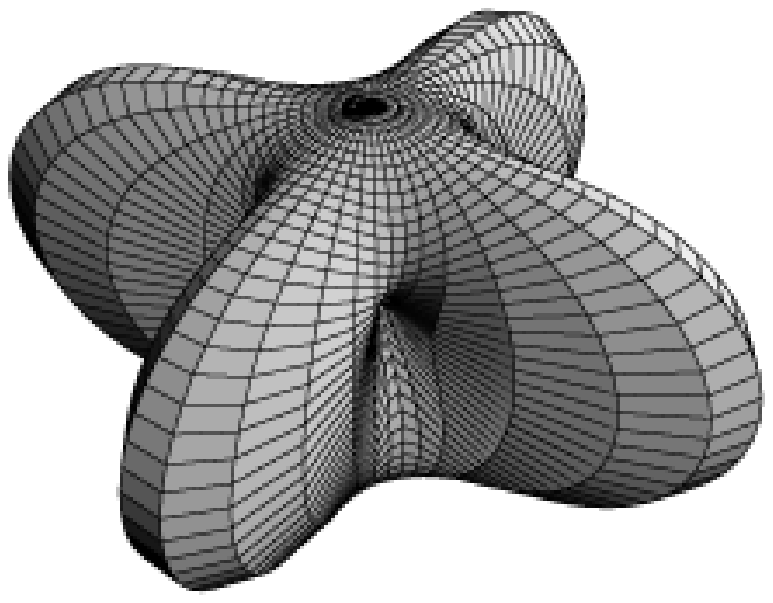}
\includegraphics[width=4cm]{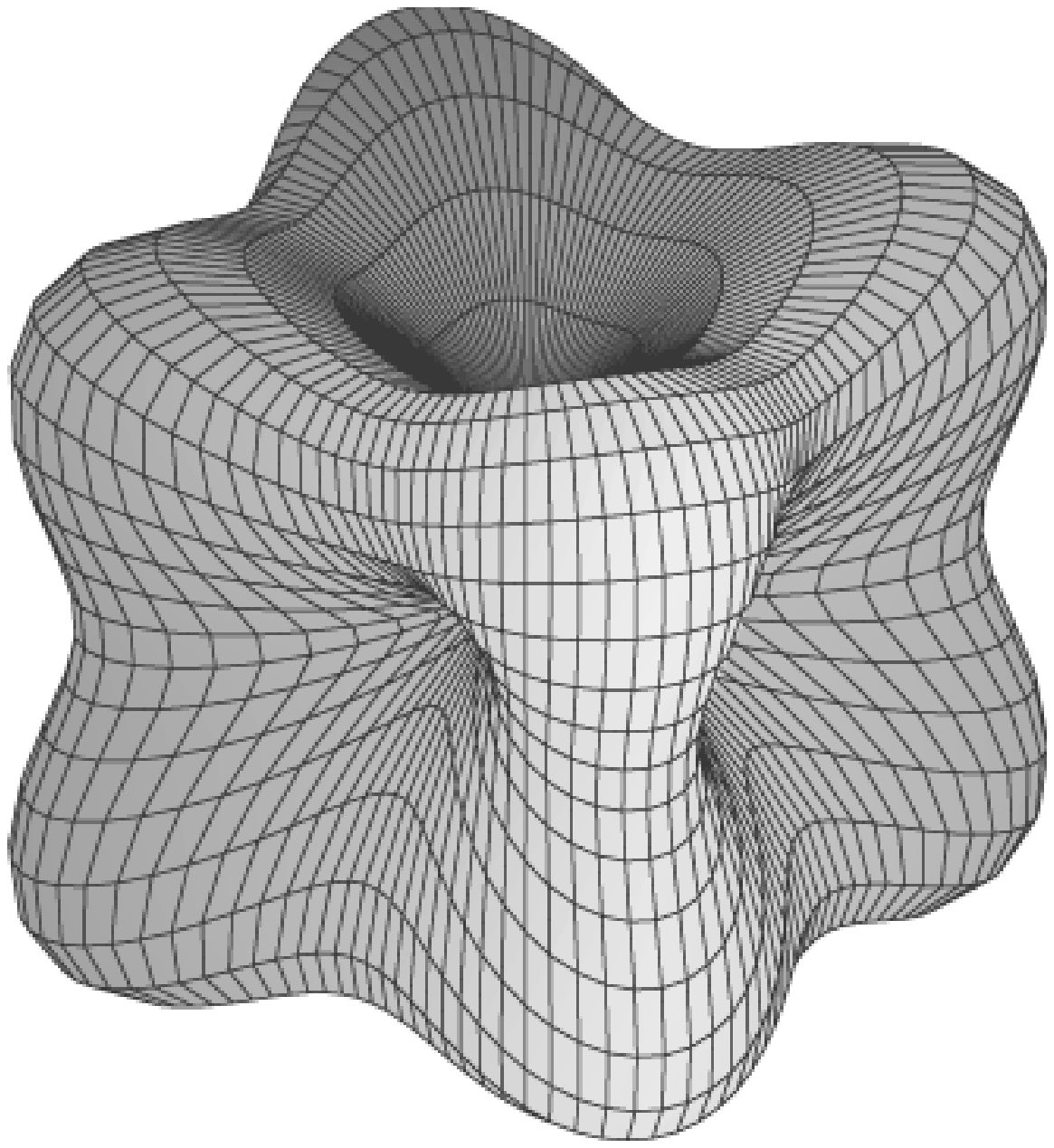}
\includegraphics[width=5.3cm]{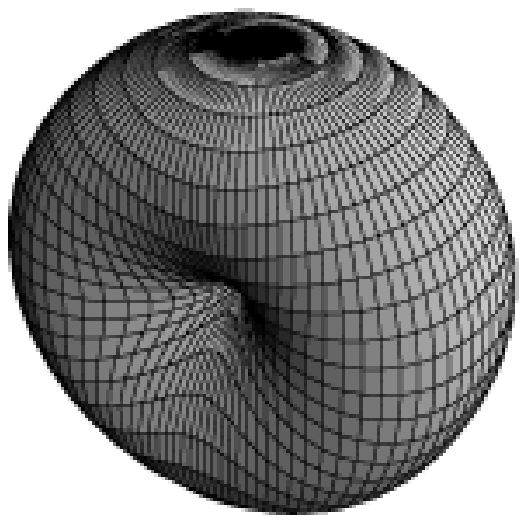}
\caption{\label{gap-functions} From top left: 2D $f$-wave in Sr$_2$RuO$_4$, $d_{x^2-y^2}$-wave in CeCoIn$_5$ and
in $\kappa$-(ET)$_2$Cu(NCS)$_2$, $(s+g)$-wave in YNi$_2$B$_2$C, $(p+h)$-wave in the A phase of PrOs$_4$Sb$_{12}$,
$(p+h)$-wave in the B phase of PrOs$_4$Sb$_{12}$.}
\end{figure}

In a seminal paper Volovik [\onlinecite{volovik}] has shown that the quasiparticle density of states in the
vortex state of the $d$-wave superconductors is calculable within a semiclassical approximation. The
predicted $\sqrt{H}$ dependence of the specific heat has been confirmed experimentally in single
crystals of YBCO [\onlinecite{moler,revaz}], LSCO [\onlinecite{chen}]
and Sr$_2$RuO$_4$ [\onlinecite{nishizaki,won-3}]. This semiclassical approach has been extended in a variety of
directions [\onlinecite{kubert-1,kubert-2,vehkter,won-4,dahm,won-5}]. When high-quality single crystals in the
extremely clean limit (i.e. $l\gg\xi$ where $l$ is
the quasiparticle mean free path and $\xi$ the superconducting coherence length) are available, the angle-
dependent thermal conductivity in the vortex state provides unique access to the gap symmetry
$\Delta(\mathbf{k})$. In the last few years Izawa et al. have determined $\Delta(\mathbf{k})$ in
Sr$_2$RuO$_4$ [\onlinecite{izawa-1}], CeCoIn$_5$ [\onlinecite{izawa-2}], $\kappa$-(ET)$_2$Cu(NCS)$_2$
[\onlinecite{izawa-3}], YNi$_2$B$_2$C [\onlinecite{izawa-4}] and PrOs$_4$Sb$_{12}$
[\onlinecite{izawa-5,maki-2}]. These gap functions are shown in Fig.~\ref{gap-functions}.

In the following, we first focus on the quasiparticle spectrum in a variety of nodal superconductors.
Then the effect of impurity scattering and the universal heat conduction is briefly summarized. The
quasiclassical approximation in the vortex state in nodal superconductors is the central part of this course. Also
the properties of nodal superconductivity in YNi$_2$B$_2$C and PrOs$_4$Sb$_{12}$ are briefly summarized. In the last
chapter we discuss unconventional density wave and gossamer superconductivity, which indicate new directions
to follow in new materials.

\section{Quasiparticle spectrum in Nodal Superconductors}

Following the BCS paper [\onlinecite{landau}] we consider the effective Hamiltonian

\begin{equation}
H=\sum_{\mathbf{k},\alpha}\xi(\mathbf{k})c^+_{\mathbf{k}\alpha}c_{\mathbf{k}\alpha}
+\frac{1}{2}\sum_{\mathbf{k},\alpha,\mathbf{k'},\alpha'}v(\mathbf{k,k'})
c^+_{\mathbf{k'}\alpha'}c^+_{-\mathbf{k'},-\alpha'}c_{-\mathbf{k},-\alpha}c_{\mathbf{k}\alpha},\label{BCS-hamiltonian}
\end{equation}
with
\begin{equation}
v(\mathbf{k,k'})=-\langle|f|^2\rangle^{-1}Vf(\mathbf{k})f(\mathbf{k'}),
\end{equation}
and
\bea
\langle|f|^2\rangle &=& \frac{1}{4\pi}\int d\Omega|f(\mathbf{k})|^2 \,\,\,\,\,\,\mathrm{3D}\\
&=& \frac{1}{(2\pi)^2}\int d\chi d\phi |f(\mathbf{k})|^2\,\,\,\,\,\,\,\mathrm{2D}
\eea
depending on whether the system is 3D or quasi-2D. In the following, we consider a group of quasi-2D superconductors, whose
quasiparticle density of states, thermodynamics etc. are identical [\onlinecite{dahm}]. We define quasi
2D systems by a cylindrical Fermi surface, as in high-$T_c$ cuprates, Sr$_2$RuO$_4$, CeCoIn$_5$,
$\kappa$-(ET)$_2$Cu(NCS)$_2$ etc. Also we consider a group of nodal superconductors with
$f(\mathbf{k})=\cos(2\phi),\sin(2\phi)$ ($d$-wave),
$e^{\pm i\phi}\cos(\chi)$ ($f$-wave), $\cos\chi$ ($d$-wave), $\sin\chi$ ($p$-wave), $e^{\pm i\phi}\sin(\chi)$ ($d$-wave),
$\cos(2\chi)$ ($g$-wave) etc. Here $\chi=ck_z$. Then it can be readily shown (Exercise 1) that these superconductors
have an identical quasiparticle density of states (DOS).

Within the mean-field approximation, i.e. the BCS approximation, Eq.~\eqref{BCS-hamiltonian} is transformed as

\begin{equation}
H=\sum_{\mathbf{k},\alpha}\Psi^+_{\mathbf{k},\alpha}(\xi(\mathbf{k})\rho_3+\Delta(\mathbf{k})\rho_1)
\Psi_{\mathbf{k},\alpha}-\sum_{\mathbf{k}}\frac{|\Delta(\mathbf{k})|^2}{v}
\end{equation}
The corresponding Nambu-Gor'kov Green function [\onlinecite{gorkov-1,nambu}] is given by

\begin{equation}
G^{-1}(\mathbf{k},\omega)=\omega-\xi(\mathbf{k})\rho_3-\Delta(\mathbf{k})\rho_1,\label{green}
\end{equation}
where the $\rho_i$'s are Pauli matrices operating on the Nambu spinor space. For simplicity we consider here only 
spin singlet pairing and $f$ as a real function.

Then the poles of the Green function Eq.~\eqref{green} give the quasiparticle energy

\begin{equation}
\omega=\pm\sqrt{\xi^2(\mathbf{k})+\Delta^2(\mathbf{k})}\simeq\pm\sqrt{v^2(k_\parallel-k_F)^2+v_2^2k_\perp^2}.
\label{poles}
\end{equation}
The last expression is an approximation near the Dirac cone.

\begin{figure}[t!]
\includegraphics[width=7cm]{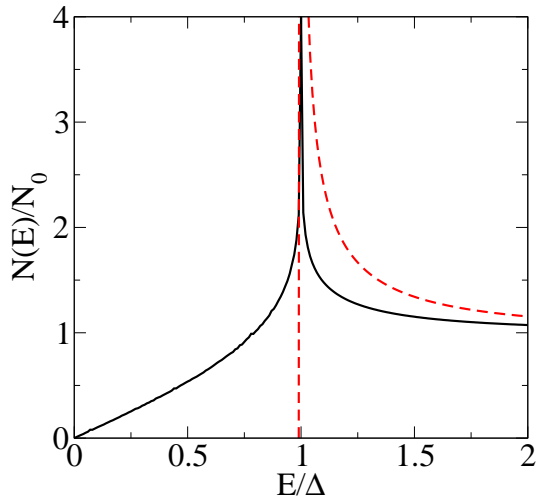}
\caption{\label{qp-dos} {The quasiparticle density of states for nodal superconductors (solid line) and for
$s$-wave superconductors (dashed line).}}
\end{figure}

From Eq.~\eqref{poles} the quasiparticle density of states is obtained as [\onlinecite{won-1}]

\begin{equation}
g(E)=\text{Re}\left\langle\frac{|E|}{\sqrt{E^2-\Delta^2|f|^2}}\right\rangle=
\left\{
\begin{array}{ll}
\frac{2}{\pi}|x|K(x) & \text{for $|x|<1$,}\\[10pt]
\frac{2}{\pi}K(1/x) & \text{for $|x|>1$,}
\end{array}
\right.\label{qp-DOS}
\end{equation}
where $x=|E|/\Delta$ and $K(k)$ is the complete elliptic integral of the first kind. We show in Fig.~\ref{qp-dos}
the quasiparticle density of states versus $x$, compared with the one for $s$-wave superconductor
with a full energy gap.

For small energies, the density of states can be expanded as
\begin{equation}
g(E)\simeq|E|/\Delta
\end{equation}
for $|E|/\Delta\ll 1$.

The mean-field approximation also gives the gap equation

\begin{equation}
\begin{split}
\lambda^{-1}&=2\pi T\langle|f|^2\rangle^{-1}\sideset{}{'}\sum_n\left\langle\frac{f^2}{\sqrt{\omega_n^2+f^2}}\right
\rangle\\
&=\langle|f|^2\rangle^{-1}\int_0^{E_0}dE\,\text{Re}\left\langle\frac{|E|}{\sqrt{E^2-\Delta^2|f|^2}}\right\rangle
\tanh(\frac{E}{2T}),\label{gap-equation}
\end{split}
\end{equation}
which we have written in 2 alternative forms. Here $\lambda=\nu N_0$ 
is the dimensionless coupling constant,
$\omega_n$ is the Matsubara frequency and $E_0$ is the cut-off energy. Also the $\omega_n$ sum in the first equation
has to be cut off at $\omega_n\simeq E_0$.

\begin{figure}[t!]
\includegraphics[width=8.75cm]{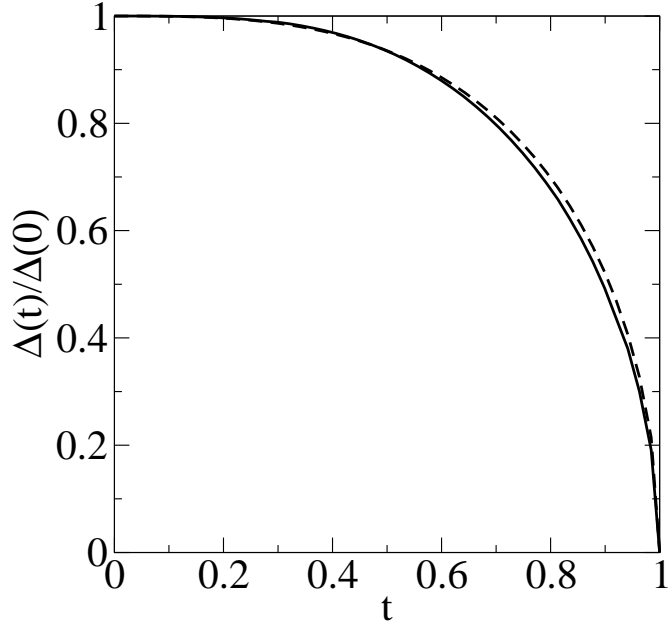}
\caption{\label{order-parameter} The temperature dependence of the order parameter (dashed - approximate, solid - exact}
\end{figure}

Then within the weak coupling limit we find [\onlinecite{dora-2}]

\begin{equation}
\Delta(0)/T_c=2.14,
\end{equation}
\begin{equation}
\Delta(t)/\Delta(0)\simeq\sqrt{1-t^3},
\end{equation}
where $t=T/T_c$. In Fig.~\ref{order-parameter} we show $\Delta(t)/\Delta(0)$ versus $t$ together with the approximate
expression Eq.(18).

As to the thermodynamics, it is convenient to start with the entropy $S$ given by

\begin{equation}
\begin{split}
S&=-4N_0\int_0^\infty dE\,g(E)[f\ln f+(1-f)\ln(1-f)]\\
&=4N_0\int_0^\infty dE\,g(E)[\ln(1+e^{-\beta E})+\beta E(1+e^{\beta E})].
\end{split}
\end{equation}
Here $N_0$ is the quasiparticle density of states at the Fermi surface in the normal state and
$f=(1+e^{\beta E})^{-1}$ is the Fermi function.

From $S$, the specific heat and the thermodynamic critical field are obtained by

\begin{equation}
C_s/T=\frac{\partial S}{\partial T},
\end{equation}
and

\begin{equation}
\frac{1}{8\pi}H_c^2(T)=F_n(T)-F_s(T)=\int_T^{T_c}dT'\,(S_n(T')-S_s(T')).
\end{equation}
Using the density of states g(E), we obtain

\begin{equation}
\frac{C_s}{\gamma_sT}=\frac{27\zeta(3)}{2\pi^2}\frac{T}{\Delta}\quad\quad\text{for $T\ll\Delta$},
\end{equation}
and

\begin{equation}
H_c(0)=\sqrt{2\pi N_0}\Delta(0).
\end{equation}

Similarly the superfluid density within the $a-b$ plane is given by

\begin{equation}
\rho_s(T)/\rho_s(0)=2\pi T\sum_n\left\langle\frac{\Delta^2f^2}{(\omega_n^2+\Delta^2f^2)^{3/2}}\right\rangle.
\end{equation}
In the limit $T\to0$ this reduces to

\begin{equation}
\rho_s(T)/\rho_s(0)=1-2\ln2\frac{T}{\Delta(0)}.
\end{equation}
In other words one can obtain $\Delta(0)$ from the $T$ linear slope of the superfluid density.

There is still a controversy as to the correct expression of the $c$ axis superfluid density. The simplest
assumptions [\onlinecite{maki-3}] give

\begin{equation}
\begin{split}
\rho_{s,c}(T)/\rho_{s,c}(0)&=\frac{\pi}{2}\frac{\Delta(t)}{\Delta(0)}\langle f\tanh(\Delta f/(2T))\rangle\\
&\simeq 1-\frac{\pi^2}{6}\left(\frac{T}{\Delta(0)}\right)^2.
\end{split}
\end{equation}

Finally, the spin susceptibility and the nuclear spin lattice relaxation rate are given by

\begin{equation}
\chi_s/\chi_n=1-\rho_s(T)/\rho_s(0),
\end{equation}
and

\begin{equation}
\begin{split}
T_1^{-1}/T_{1n}^{-1}&=\int_0^\infty\frac{dE}{2T}\,g^2(E)\text{sech}^2(\frac{E}{2T})\\
&\simeq\frac{\pi^2}{3}\left(\frac{T}{\Delta}\right)^2.
\end{split}
\end{equation}

We stress again that these expressions are not only valid for $d$-wave superconductors in the weak-coupling limit
as in high-$T_c$ cuprates, but for all nodal superconductors with $f(\mathbf{k})$ given above. Therefore in order
to explore the individual gap symmetry, we have to look at other properties.

It is somewhat surprising that all the energy gaps of the identified nodal superconductors in the quasi-2D systems belong to
the above class of $f$'s.

\vskip 1cm
\noindent
\textbf{Exercises 1.}

\begin{enumerate}
\item[1.1] Within the weak-coupling theory calculate the jump in the specific heat at $T=T_c$

\begin{enumerate}
\item[a.] for $s$-wave superconductivity. (Answer: $\Delta C/C_n=12/(2\zeta(3))=1.43$)
\item[b.] for nodal superconductivity with $\Delta(\mathbf{k})=\Delta f(\mathbf{k})$.
(Answer: $\Delta C/C_n=12/(7\zeta(3))\langle|f|^4\rangle/\langle|f|^2\rangle^2$) 
\end{enumerate}

\item[1.2] Evaluate the ratio $\Delta(0)/T_c$ within the weak-coupling theory

\begin{enumerate}
\item[a.] for $s$-wave superconductivity. (Answer: $\Delta(0)/T_c=\pi/\gamma=1.76$)
\item[b.] for nodal superconductivity. (Answer: $\Delta(0)/T_c=(\pi/\gamma)\exp(-\langle f^2\rangle^{-1}
\langle f^2\ln f\rangle)$)
\end{enumerate}

\item[1.3] Show that the quasiparticle density of states for a group of $f$'s discussed is the same as given by
Eq.~\eqref{qp-DOS}.

\item[1.4] Express $\text{Re}\langle f^2/\sqrt{x^2-f^2}\rangle$ in terms of complete elliptic integrals. Answer:
\begin{equation}
\text{Re}\left\langle\frac{f^2}{\sqrt{x^2-f^2}}\right\rangle=
\left\{
\begin{array}{ll}
\frac{2}{\pi}(K(x)-E(x)) & \text{for $x<1$,}\\[10pt]
\frac{2}{\pi}x(K(1/x)-E(1/x)) & \text{for $x>1$.}
\end{array}
\right.
\end{equation}
\end{enumerate}

\section{Effect of Impurity Scattering}

In metals the presence of impurities or foreign atoms is unavoidable. Also, they provide the simplest agents of
quasiparticle relaxation. Therefore the study of impurity scattering is crucial to understand quasiparticle
transport such as electric conductivity and thermal conductivity.

In the early sixties the effect of impurity scattering in $s$-wave superconductivity was systematically studied
in [\onlinecite{anderson-2,abrikosov-1,abrikosov-2}]. As is well known, nonmagnetic impurity scattering has
little effect in $s$-wave superconductors.
The superconducting transition temperature and the thermodynamics are almost unaffected.  The most dominant effect is
a reduction of the quasiparticle mean-free path, as in the normal state, and its consequence on the magnetic
penetration depth. On the other hand, magnetic impurities have a profound effect on $s$-wave superconductivity.
The superconducting transition temperature is sharply reduced. Also in some cases gapless superconductivity is
induced [\onlinecite{abrikosov-2,maki-4}].

In contrast, the nonmagnetic impurities have profound effects on nodal superconductors as pointed out in
[\onlinecite{pethick,schmitt,hirschfeld-1}].
In particular, resonant impurity scattering appears to be prevalent. The extreme limit is the unitary limit
where a resonance occurs at $E=0$. Such a model has been discussed for $d$-wave superconductivity in
high-$T_c$ cuprates in [\onlinecite{hirschfeld-2,hotta}]. The first self-consistent studies of impurity scattering
in $d$-wave superconductors were performed in [\onlinecite{sun-1,sun-2,maki-1}].

For simplicity we shall limit ourselves to the unitary limit. Also we assume that the impurity
is point like, i.e. it has only $s$-wave scattering amplitude. Then the effect of impurity scattering can be
incorporated by $\omega\to\tilde\omega$ in Eq.~\eqref{green} where $\tilde\omega$ is the renormalized frequency
given by [\onlinecite{sun-1}]

\begin{equation}
\tilde\omega=\omega-\frac{\pi}{2}\Gamma\frac{\sqrt{1-\tilde x^2}}{\tilde x}\left(K\left(\frac{1}{\sqrt{1-\tilde x^2}}
\right)\right)^{-1},\label{renorm-freq}
\end{equation}
and $\tilde x=\tilde\omega/\Delta$ and $\Gamma=n_i(\pi N_0)^{-1}$ is the quasi-particle scattering rate in the
normal state. Then the quasiparticle density of states in the presence of impurities is given by

\begin{equation}
g(E,\Gamma)=\frac{2}{\pi}\text{Re}\left\{\frac{\tilde x}{\sqrt{1-\tilde x^2}}K\left(\frac{1}{\sqrt{1-\tilde x^2}}
\right)\right\}.
\end{equation}

\begin{figure}[t!]
\psfrag{x}[t][b]{$E/\Delta$}
\psfrag{y}[b][t]{$g(E,\Gamma)/N_0$}
\includegraphics[width=10cm]{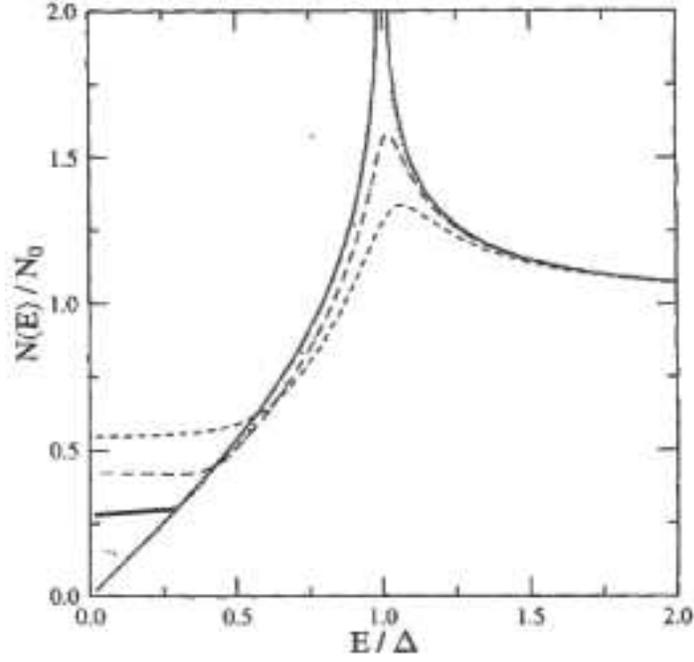}
\caption{\label{impurity-dos} The quasiparticle density of states in the presence of impurities, for
several impurity concentrations.}
\end{figure}

The quasiparticle density of states in the presence of impurities is shown in Fig.~\ref{impurity-dos}. We note a
rapid appearance of disorder-induced spectral weight at $E=0$. Indeed $g(0,\Gamma)$ is given by

\begin{equation}
g(0,\Gamma)=\frac{C_0}{\sqrt{1+C_0^2}}K\left(\frac{1}{\sqrt{1+C_0^2}}\right),\label{31}
\end{equation}
where $C_0$ is obtained from

\begin{equation}
\frac{C_0^2}{\sqrt{1+C_0^2}}=\frac{\pi}{2}\frac{\Gamma}{\Delta}K^{-1}\left(\frac{1}{\sqrt{1+C_0^2}}\right).\label{32}
\end{equation}
In the limit $\Gamma/\Delta\to0$, both Eq.~\eqref{31} and Eq.~\eqref{32} reduce to

\begin{equation}
g(0,\Gamma)=C_0\ln(\frac{4}{C_0})\simeq\left(\frac{\pi\Gamma}{2\Delta}\ln^{-1}
\left(4\sqrt{\frac{2\Delta}{\pi\Gamma}}\right)\right)^{1/2},\label{33}
\end{equation}
and

\begin{equation}
C_0^2\ln(4/C_0)\simeq\frac{\pi\Gamma}{2\Delta}.\label{34}
\end{equation}

The gap equation in the presence of impurity scattering is given by

\begin{equation}
\begin{split}
\lambda^{-1}&=2\pi T\langle f^2\rangle^{-1}\sideset{}{'}
\sum_n\left\langle\frac{f^2}{\sqrt{\tilde\omega_n^2+\Delta^2f^2}}\right\rangle\\
&=\frac{8T}{\Delta}\sideset{}{'}\sum_n\sqrt{1+\tilde x_n^2}\left(E\left(\frac{1}{\sqrt{1+\tilde x_n^2}}\right)
-\frac{\tilde x_n^2}
{1+\tilde x_n^2}K\left(\frac{1}{\sqrt{1+\tilde x_n^2}}\right)\right),\label{impurity-gap}
\end{split}
\end{equation}
where $\tilde x_n=\tilde\omega_n/\Delta$ and $\tilde\omega_n$ is the renormalized Matsubara frequency. Then in the
limit $\Delta\to0$, we obtain the Abrikosov-Gor'kov formula:

\begin{equation}
-\ln\left(\frac{T_c}{T_{c0}}\right)=\Psi\left(\frac{1}{2}+\frac{\Gamma}{2\pi T_c}\right)
-\Psi\left(\frac{1}{2}\right),\label{abrikosov}
\end{equation}
where $T_c\,(T_{c0})$ is the superconducting transition temperature in the presence (absence) of impurities.
Here $\Psi(z)$ is the digamma function.

Note that Eq.~\eqref{abrikosov} is the same as for $s$-wave superconductors in the presence of magnetic
impurities [\onlinecite{abrikosov-2}]. In nodal superconductors $\Gamma$ is due to nonmagnetic impurities, 
whereas in $s$-wave superconductors $\Gamma$ is associated with magnetic 
scattering which involves spin flipping.

\begin{figure}[t!]
\psfrag{x}[t][b]{$\Gamma/\Gamma_c$}
\psfrag{y}[b][t]{$\Delta(0,\Gamma)/\Delta_{00}$, $T_c/T_{c0}$ \text{and} $g(0,\Gamma)/N_0$}
\includegraphics[width=10cm]{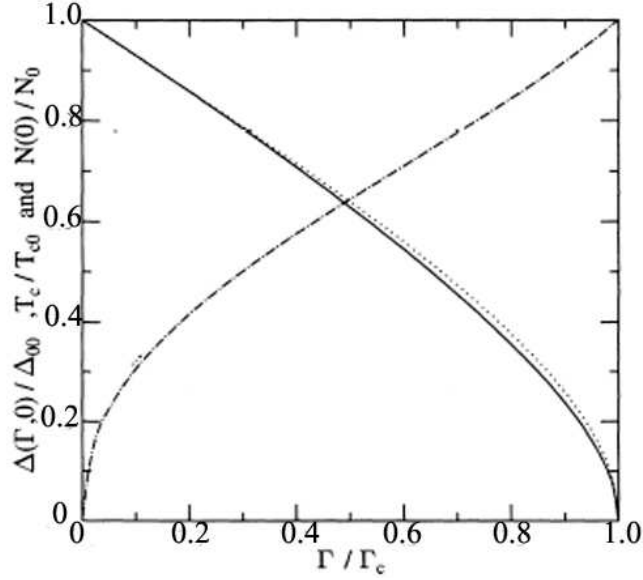}
\caption{\label{delta-gamma} $\Delta(0,\Gamma)/\Delta_{00}$ (dashed line), $T_c/T_{c0}$ (solid line)
and $g(0,\Gamma)/N_0$ (dashed-dotted line) are shown as a function of $\Gamma/\Gamma_c$ in the unitary limit.}
\end{figure}

At $T=0$K Eq.~\eqref{impurity-gap} reduces to

\begin{equation}
-\ln\left(\frac{\Delta(\Gamma,0)}{\Delta(0,0)}\right)=2\left\langle f^2\ln(C_0+\sqrt{C_0^2+f^2})\right\rangle
-\frac{2\Gamma}{\Delta}\int_{C_0}^\infty dx\,x^2(1-E/K)[(1+x^2)E-K],\label{gap-imp}
\end{equation}
where $E=E(1/\sqrt{1+x^2})$ and $K=K(1/\sqrt{1+x^2})$ and $C_0$ has already been defined in 
Eq.~\eqref{32}.
We show in Fig.~\ref{delta-gamma} $T_c/T_{c0},\ \Delta(0,\Gamma)/\Delta(0,0,)$ and $g(0)$ versus $\Gamma/\Gamma_c$ where
$\Gamma_c=\pi T_{c0}/(2\gamma)\simeq 0.8819 T_{c0}$.

\begin{figure}[h!]
\includegraphics[width=9.5cm]{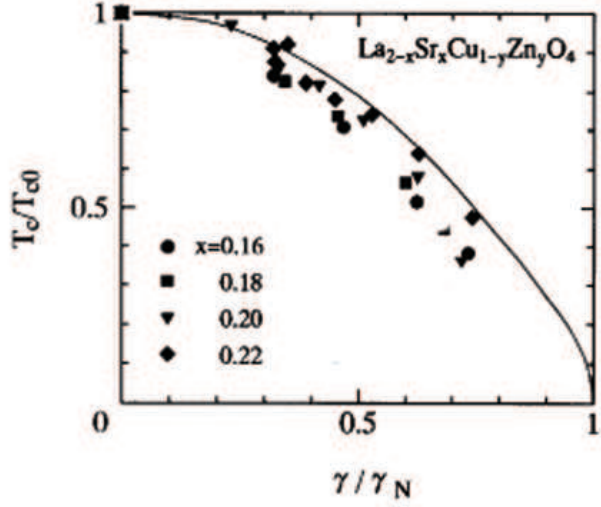}
\caption{\label{Tc-g(0)} Experimental values for $T_{c}(\Gamma)$ for doped LSCO}
\end{figure}

Surprisingly, $\Delta(0,\Gamma)/\Delta(0,0)$ follows very closely $T_c/T_{c0}$. Also $g(0,\Gamma)$ increases very
rapidly with $\Gamma$. This rapid increase in the DOS has been measured by the low temperature specific heat in
doped LSCO [\onlinecite{momono}]. Note $C_s/(\gamma_sT)=g(0)$. As seen from FIG.~\ref{Tc-g(0)} the experimental data agree very
well with the theoretical prediction. Especially the agreement is almost perfect in the vicinity of the optimally
doped LSCO. This suggests strongly that the weak-coupling BCS theory for $d$-wave superconductivity is adequate
for LSCO. We have already mentioned that $\Delta(0)/T_c=2.14$ in the weak-coupling limit. We can deduce this ratio
for optimally doped LSCO, YBCO and Bi2212. These are 2.15, 2.64 and 5.3 respectively. Indeed LSCO appears to
be in the weak-coupling limit whereas YBCO may be in the intermediate regime.
Since the large ratio 5.3 for Bi2212 is rather strange, we wonder if another order parameter is hidden to make this
large energy gap possible (see Section VIII).

\begin{figure}[b!]
\includegraphics[width=9cm]{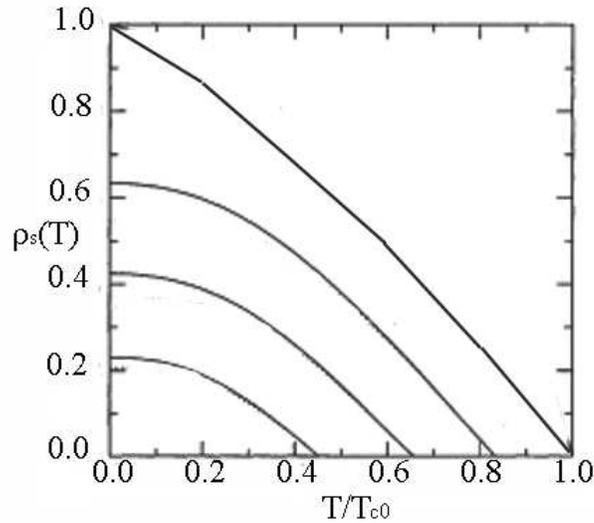}
\caption{\label{superfluid} The superfluid density versus temperature for various impurity concentrations.}
\end{figure}

The planar superfluid density in the presence of impurities is given by

\begin{equation}
\frac{\rho_s(T,\Gamma)}{\rho_s(0,0)}=2\pi T\sum_{n=0}^\infty\left\langle
\frac{\Delta^2f^2}{(\tilde\omega_n^2+\Delta^2f^2)^{3/2}}\right\rangle.
\end{equation}
For $T=0$ this reduces to

\begin{equation}
\frac{\rho_s(T,\Gamma)}{\rho_s(0,0)}=1-\frac{\Gamma}{\Delta C_0}+\frac{\Gamma}{\Delta}
\int_{C_0}^\infty\frac{dx}{x^2}\,\left(1-\frac{E}{K}\right)^2.
\end{equation}
This sharp decrease in superfluid density in Zn-substituted YBCO has been observed by the $\mu$-SR
experiments [\onlinecite{berhard}]. In Fig.~\ref{superfluid} we show $\rho_s(T,\Gamma)/\rho_s(0,0)$ versus $T/T_{c0}$.

\vskip 1cm
\noindent
\textbf{Exercises 2.}
\begin{enumerate}
\item[2.1.] Derive Eq.~\eqref{renorm-freq}.
\item[2.2.] Derive Eq.~\eqref{33} and Eq.~\eqref{34}.
\end{enumerate}

\section{Universal Heat Conduction}

\begin{figure}[t!]
\includegraphics[width=8cm]{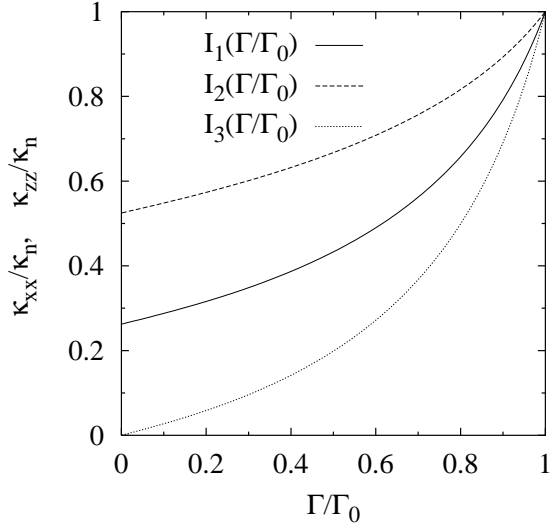}
\caption{\label{kappa} The functions $I_{1}(\Gamma/\Gamma_{c})$, 
$I_{2}(\Gamma/\Gamma_{c})$,  $I_{3}(\Gamma/\Gamma_{c})$}
\end{figure}

In 1993, Patrick Lee [\onlinecite{lee-1}] pointed out that the thermal conductivity in $d$-wave superconductors is linear
in $T$ for $T\ll\Delta(0)$, and that it takes the universal value $\kappa_{00}/T=k_B^2vn/(3\hbar v_2)$ in the
limit $\Gamma\to0$. A further study indicates that $\kappa/T$ increases very rapidly with $\Gamma$, the
quasiparticle scattering due to impurities [\onlinecite{sun-2}]. This result was used by May Chiao et al
[\onlinecite{chiao-1,chiao-2}]
to extract $\Delta(0)/E_F$ of optimally doped Bi2212 and YBCO. By transforming the expression of the
thermal conductivity given by Ambegaokar and Griffin [\onlinecite{ambegaokar}], the thermal conductivity
for $T\ll\Delta(0)$ is given by [\onlinecite{sun-2}]

\begin{equation}
\begin{split}
\kappa_{xx}/\kappa_n=\kappa_{yy}/\kappa_n &=\frac{\Gamma}{\Delta}
\left\langle\frac{C_0^2}{(C_0^2+f^2)^{3/2}}\right\rangle\\
&=\frac{2\Gamma}{\pi\Delta}\frac{1}{\sqrt{1+C_0^2}}E\left(\frac{1}{\sqrt{1+C_0^2}}\right),\label{therm-cond}
\end{split}
\end{equation}
where $C_0$ and $\Delta=\Delta(0,\Gamma)$ have been given in Eq.~\eqref{32} and Eq.~\eqref{gap-imp}
respectively and $E(k)$ is the complete elliptic integral of the second kind. For later purposes, it is more
convenient to normalize $\kappa_{xx}$ by $\kappa_n^c$, the normal state thermal conductivity, by
$\Gamma=\Gamma_c=0.882T_{c0}$ the critical scattering where superconductivity disappears.

\begin{equation}
\kappa_n^c=\frac{\pi^2Tn}{6\Gamma_cm}.
\end{equation}
Then we can rewrite Eq.~\eqref{therm-cond} as

\begin{equation}
\kappa_{xx}/\kappa_n^c=\frac{2\Gamma_c}{\pi\Delta}\frac{1}{\sqrt{1+C_0^2}}E\left(\frac{1}{\sqrt{1+C_0^2}}\right)
=I_1(\Gamma/\Gamma_c).
\end{equation}
$I_1(\Gamma/\Gamma_c)$ versus $\Gamma/\Gamma_c$ is shown in Fig.~\ref{kappa}.

\begin{figure}[t!]
\includegraphics[width=8cm]{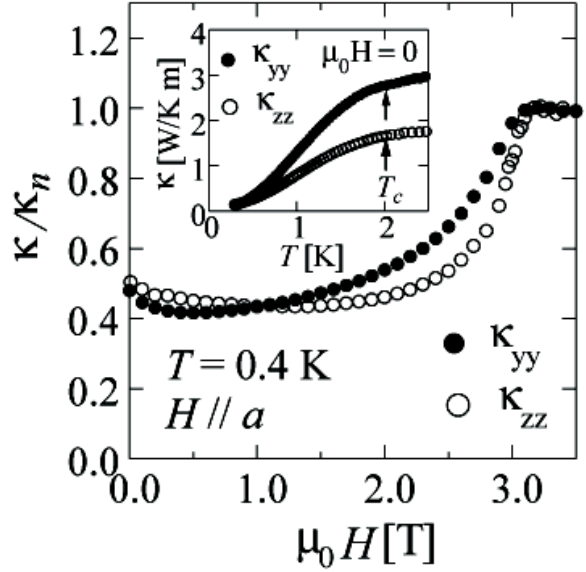}
\caption{\label{kappa-data} Thermal conductivity data for UPd$_{2}$Al$_{3}$}
\end{figure}

We note that $\kappa_{xx}$ increases monotonically with $\Gamma$. In other words, the thermal conductivity
increases with the impurity scattering. This counterintuitive behavior is understood, if one realizes
that the impurity scattering produces quasiparticles due to the pair-breaking effect [\onlinecite{sun-2}].
Indeed, the predicted $\Gamma$ dependence of $\kappa_{xx}$ was verified in Zn-doped YBCO [\onlinecite{taillefer}]
and more recently in Sr$_2$RuO$_4$ [\onlinecite{suzuki}]. If we substitute in Eq.~\eqref{therm-cond}
$f=\sin(2\phi)$, $\cos\chi$, $e^{\pm i\phi}\cos\chi$, $\sin\chi$, $e^{\pm i\phi}\sin\chi$ or $\cos(2\chi)$, we
will obtain the same result. The planar thermal conductivity can thus not 
discriminate between different nodal superconductors
[\onlinecite{won-new}]. The result for the out-of-plane thermal conductivity is of more interest. We obtain

\begin{equation}
\begin{split}
\kappa_{zz}/\kappa_n^c&=\frac{\Gamma_c}{\Delta}\left\langle(1-\cos(2\chi))\frac{C_0^2}{(C_0^2+|f|^2)^{3/2}}
\right\rangle\\
&=I_1(\Gamma/\Gamma_c)
\end{split}
\end{equation}
for $f=\cos(2\phi)$, $\sin(2\phi)$ and $\cos(2\chi)$. But

\begin{equation}
\begin{split}
\kappa_{zz}/\kappa_n^c&=\frac{4\Gamma_c}{\pi\Delta}\frac{1}{\sqrt{1+C_0^2}}\left((1+C_0^2)
E\left(\frac{1}{\sqrt{1+C_0^2}}\right)-K\left(\frac{1}{\sqrt{1+C_0^2}}\right)\right)\\
&=I_2(\Gamma/\Gamma_c)
\end{split}
\end{equation}
for $f=\cos\chi$, $e^{\pm i\phi}\cos\chi$ and

\begin{equation}
\begin{split}
\kappa_{zz}/\kappa_n^c&=\frac{4\Gamma_c}{\pi\Delta}\frac{C_0^2}{\sqrt{1+C_0^2}}\left(
K\left(\frac{1}{\sqrt{1+C_0^2}}\right)-E\left(\frac{1}{\sqrt{1+C_0^2}}\right)\right)\\
&=I_3(\Gamma/\Gamma_c)
\end{split}
\end{equation}
for $f=\sin\chi$ and $e^{\pm i\phi}\sin\chi$. We show $I_1(\Gamma/\Gamma_c)$, $I_2(\Gamma/\Gamma_c)$ and
$I_3(\Gamma/\Gamma_c)$ versus $\Gamma/\Gamma_c$ in FIG.~\ref{kappa}. It is clear that $\kappa_{zz}\to0$
for $f=\sin\phi$ and $e^{\pm i\phi}\sin\phi$ in the limit $\Gamma\to0$. There is no universal heat conduction.
Very recently thermal conductivity data for $\kappa_{xx}$ and $\kappa_{zz}$ at $T=0.4$K and
$\mathbf{H}\parallel\mathbf{\hat z}$ in UPd$_2$Al$_3$ was reported [\onlinecite{won-new,watanabe}].
This is shown in Fig.~\ref{kappa-data}.
Although the field dependence of $\kappa_{ii}$ is not the same as its $\Gamma$ dependence, we conclude
$\Delta(\mathbf{k})$ in UPd$_2$Al$_3$ is more consistent with $\Delta(\mathbf{k})\sim\cos(2\chi)$ rather
than $\Delta(\mathbf{k})\sim\cos\chi$. The latter has been proposed in 
[\onlinecite{machale}] based on antiparamagnon
exchange with $\mathbf{Q}=(0,0,\pi/c)$ [\onlinecite{bernhoeft}].

\vskip 0.5cm
\noindent
\textbf{Exercises 3.}
\begin{enumerate}
\item[3.1.] Derive Eq.~\eqref{therm-cond} for a variety of $f$'s.
\item[3.2.] Calculate $\kappa_{zz}$ for these $f$'s.
\end{enumerate}

\section{Quasiclassical approximation}

\subsection{Vortex state}

In the presence of a magnetic field, all known nodal superconductors enter into the vortex state when the magnetic
field exceeds $H_{c1}(T)$, the lower critical field. In this vortex state the quantized vortex lines form a
regular two dimensional lattice in the equilibrium configuration. Also each vortex line carries a unit flux
$\phi_0=hc/(2e)=2.06\cdot10^{-7}$Gcm$^2$. The marvelous theory of type II superconductivity was created by
Abrikosov [93] in 1957, based on the phenomenological Ginzburg-Landau theory [\onlinecite{ginzburg}].
The microscopic foundation of GL theory was provided by Gor'kov [\onlinecite{gorkov-2}] after the appearance of
the BCS theory [\onlinecite{bardeen}].
In $s$-wave superconductors the vortex lattice is usually hexagonal. But in $d$-wave superconductors the
square vortex lattice is more stable for $T/T_c\le0.9$ [\onlinecite{won-6,shiraishi}].
Indeed the square vortex lattice was recently observed in LSCO by neutron scattering experiments [\onlinecite{gilardi}].
Also unlike in $s$-wave superconductors, the quasiparticles dominate the low-temperature transport properties
in the whole vortex state in nodal superconductors.  The quasiclassical approximation then provides
the most practical way to handle the quasiparticle transport in the vortex state for $T \ll \Delta$.

\subsection{Quasiparticle spectrum in the vortex state}

In 1993 Volovik [\onlinecite{volovik}] showed how to calculate the quasiparticle DOS in the vortex state
of $d$-wave superconductors,when $T\ll\Delta(0)$. Here we shall follow this procedure. First we note that the
quasiparticle energy $E_\mathbf{k}$ is shifted to $E_\mathbf{k}-\mathbf{v \cdot q}=E_\mathbf{k}-\mathbf{k \cdot v}_s$ in the
presence of a superflow [\onlinecite{maki-5}]. Here $\mathbf{v}$, $2\mathbf{q}$ and $\mathbf{v}_s$
are the quasiparticle velocity, the pair momentum and the superfluid velocity respectively. Also for a class of
nodal superconductors we considered in Section II, we obtain $g(E)=|E|/\Delta$ for $|E|\ll\Delta$.

In the presence of a superflow this is generalized as
\begin{equation}
g(E,\mathbf{H})=\Delta^{-1}\langle|E-\mathbf{v \cdot q}|\rangle,
\end{equation}
or
\begin{equation}
g(0,\mathbf{H})\equiv g(\mathbf{H})=\Delta^{-1}\langle|\mathbf{v \cdot q}|\rangle,\label{dos-magnetic}
\end{equation}
where $\langle\dots\rangle$ denotes the averages over the Fermi surface and over the vortex lattice. Here
$\mathbf{v \cdot q}$ is called the Doppler shift.

Let us consider $d$-wave superconductors in a magnetic field $\mathbf{H}\parallel\mathbf{\hat c}$. Then
Eq.~\eqref{dos-magnetic} becomes

\begin{equation}
\begin{split}
g(\mathbf{H})&=\frac{4v}{\pi\Delta d^2}\int_0^d\frac{r\,dr}{2\pi}\int_0^{\pi/2}d\alpha\,\cos\alpha
\int_0^{2\pi}\frac{d\phi}{2\pi}\,\delta(\cos(2\phi))\\
&=\frac{2}{\pi^2}\frac{v\sqrt{eH}}{\Delta}.\label{dos-2}
\end{split}
\end{equation}
Here $d=1/\sqrt{eH}$ and we assumed for simplicity a square vortex lattice with lattice constant $d$. Also,
in the present calculation we took $\mathbf{q}=\hat{\phi}/(2r)$, where $r$ is the distance from the center of a
vortex and $\alpha$ is the angle between $\mathbf{v}$ and $\mathbf{q}$. We treated the average over the vortex
lattice  \`a la Wigner-Seitz. Finally in the earlier treatments
[\onlinecite{kubert-1,kubert-2,won-4,dahm}] the factor $\pi^{-1}$ coming from
$(2\pi)^{-1}\int_{0}^{2\pi}d\phi\,\delta(\cos(2\phi))$ was missing. From Eq.~\eqref{dos-2} the
specific heat and other observables [\onlinecite{won-7}] are obtained as

\begin{gather}
C_s/\gamma_nT=g(\mathbf{H}),\quad\quad \kappa_s/\kappa_n=g(\mathbf{H}),\notag\\
\frac{\rho_s(\mathbf{H})}{\rho_s(0)}=1-g(\mathbf{H}).
\end{gather}

We have already mentioned that the $\sqrt{H}$ dependence of the specific heat in the nodal superconductors was
observed in
YBCO [\onlinecite{moler,revaz}], LSCO [\onlinecite{chen}] and Sr$_4$RuO$_4$ [\onlinecite{nishizaki,won-3}].
However,  in the earlier analysis [\onlinecite{moler,revaz}] a factor $\sim0.3$ was missing. We believe the $\pi^{-1}$
we find here accounts for this missing factor. In other words, the semiclassical result is not only qualitatively but
also quantitatively accurate. It is very easy to work out $\langle|\mathbf{v \cdot q}|\rangle$ for other classes of
$\Delta(\mathbf{k})$ as discussed in Section II. Then it is easily seen that the configuration
$\mathbf{H}\parallel\mathbf{\hat{c}}$ cannot discriminate $d$-wave superconductors from other nodal superconductors.

\subsection{Extension to the arbitrary field orientation}

\begin{figure}[t!]
\includegraphics[width=15cm]{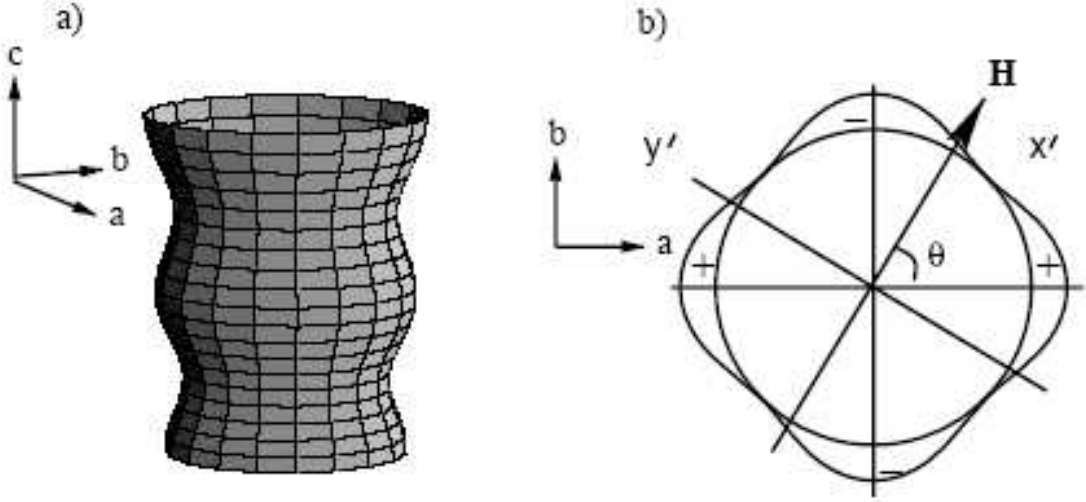}
\caption{\label{Fermi-surface} Quasi-two-dimensional cylindrical Fermi surface}
\end{figure}

In order to get a handle on the gap symmetry of $\Delta(\mathbf{k})$, it is necessary to consider the case of
arbitrary field orientations. For simplicity we limit ourselves to the case where $\mathbf{H}$ is in the $a-b$ plane
with an angle $\phi$ from the $a$ axis. Precisely this configuration is considered in [\onlinecite{vehkter}].
Unfortunately, however, a rather unrealistic Fermi surface was considered.
Instead we consider the quasi-2D Fermi surface shown in Fig.~\ref{Fermi-surface} [\onlinecite{won-8}].
In the present configuration the vortex loses circular symmetry around the vortex axis. Therefore we find it is
very useful to introduce a scale transformation as in [\onlinecite{won-9}] to make a circular vortex.
For $f=\cos(2\phi)$ we find [\onlinecite{won-4}]

\begin{equation}
\begin{split}
g(\mathbf{H})&=\frac{2}{\pi}\tilde v\sqrt{eH}\left\langle\sqrt{\sin^2\chi+\sin^2(\phi-\phi')}\right\rangle\\
&\simeq\frac{2}{\pi}\tilde v\sqrt{eH}(0.955+0.0285\cos(4\phi)),
\end{split}
\end{equation}
where $\tilde v=\sqrt{v_cv}$ and $v_c$ is the Fermi velocity parallel to the $c$ axis and $\phi$ is the angle
$\mathbf{H}$ makes from the $a$ axis. We thus find a $3\%$ fourfold term $(\sim\cos(4\phi))$ with a $\leq$.3\%
angular variation of $g({\bf H})$ The $\phi$ dependence of
the specific heat in YBCO was studied by Wang et al [\onlinecite{wang}].  They could not find this small
fourfold term which is within their experimental error.

When $\Delta(\mathbf{k})$ has a horizontal node as in Sr$_2$RuO$_4$ [\onlinecite{izawa-1}] $g(\mathbf{H})$
does not exhibit the $\phi$ dependence. In particular for $f=e^{\pm i\phi}\cos\chi$ we obtain

\begin{equation}
\langle|\mathbf{vq}|\rangle=\frac{2}{\pi^2}\tilde v\sqrt{eH}\sqrt{2}E(1/\sqrt{2})
=1.216\frac{2}{\pi^2}\tilde v\sqrt{eH}
\end{equation}
Therefore the specific heat of Sr$_2$RuO$_4$ by Deguchi et al [\onlinecite{deguchi}] appears somewhat puzzling.
Their data for $T>0.5$K is consistent with 2D $f$-wave superconductor. However, something different appears to 
happen below $T=0.3$K. From experience with YNi$_2$B$_2$C we know sharp cusps in $g(\mathbf{H})$ imply point nodes
and not line nodes [\onlinecite{maki-6}]. Therefore the specific heat for $T<0.3$K suggests the appearance
of point like minigaps for $\mathbf{k}\parallel(1,0,0)$, $(0,1,0)$ etc.  These may be reproduced with
the secondary energy gap

\begin{equation*}
\Delta_2(\mathbf{k})\sim e^{\pm i\phi}\cos\chi(1-a\cos(4\phi)\cos^2\chi)
\end{equation*}
with $a\lesssim1$. Then the secondary gap has the same symmetry as a 2D $f$-wave superconductor.

\vskip 0.5cm
\noindent
\textbf{Exercises 4.}

\begin{enumerate}
\item[4.1.] Calculate $g(\mathbf{H})$ when $\mathbf{H}$ is rotated within the $z-x$ plane. Consider here
$f=\cos\chi$, $\sin\chi$ and $\cos(2\chi)$.
\item[4.2.] Calculate $g(\mathbf{H})$ when $\mathbf{H}$ is rotated within the $x-y$ plane. Consider the same
set of $f$'s as above.
\end{enumerate}

\section{Magnetothermal conductivity}

Experiments on the angular dependent thermal conductivity in YBCO [\onlinecite{yu,aubin,ocana,106}]
indicate that the thermal conductivity is sensitive to the nodal
directions, although theoretical interpretation of this data has been given
only recently [\onlinecite{won-5}].  In order to calculate the thermal conductivity it is crucial
to incorporate the effects of impurity scattering and of the Doppler shift on an equal
footing.  Then there arise natually two limiting cases:
a) the superclean limit $(\Gamma\Delta)^{1/2} \ll <|{\bf v\cdot q}|>$ and
b) the clean limit $(\Gamma\Delta)^{1/2} \gg <|{\bf v\cdot q}|> \gg \Gamma$.

First it is necessary to determine the quasiparticle lifetime in the presence of impurity
scattering and the Doppler shift.  This is given by [\onlinecite{won-4}]
\bea
C_{0} &=& \frac{\pi\Gamma}{2\Delta}[<C_{0}\ln(\frac{2}{\sqrt{C_{0}^{2}+x^{2}}}) + x \tan^{-1}(x/C_0)>]^{-1}
\eea
where $\Delta C_{0} = Im(\tilde{\omega}) $ at $\omega=0$ is the quasi-particle relaxation rate on
the Fermi surface and $ x= |{\bf v}\cdot {\bf q}|/\Delta$.  This formula applies to the class
of f's introduced in {\bf II}, which include d-wave superconductivity with $f = \cos(2\phi)$.
Also in the later analysis we assume that ${\bf H}$ lies in the ab-plane.

\subsection{Superclean limit}

In the superclean limit we can assume that $x \gg C_{0}$.  Then Eq.(52) is solved as
\bea
C_{0} &=& \frac{\Gamma}{\Delta}<x>^{-1} - \frac{2}{\pi}(\frac{\Gamma}{\Delta})^{2}<x>^{-3}[\ln(2/x)-1]
\eea
On the other hand, in the clean limit we obtain
\bea
C_{0}^{2}\ln(2/C_{0}) &=& \frac{\pi \Gamma}{2 \Delta} - \frac{1}{2}<x^{2}> + \ldots
\eea
Also when {\bf H} is perpendicular to the a-b plane we obtain
\bea
<x^{2}>&=& \frac{1}{4\pi} \frac{v^{2}eH}{\Delta^{2}}\ln\frac{\Delta}{v\sqrt{eH}}
\eea
The thermal conductivity within the a-b plane and in the limit $T \ll \Delta$ is given by [\onlinecite{won-4}]
\bea
\kappa_{\parallel}/\kappa_{n} (\equiv \kappa_{xx}/\kappa_{n} = \kappa_{yy}/\kappa_{n})=\frac{\pi\Gamma}{4\Delta}
\left\langle \frac{1+\frac{C_{0}^{2}+x^{2}-f^{2}}{|(C_{0}+ix)^{2}+f^{2}|}}{Re\sqrt{(C_{0}+ix)^{2}+f^{2}}}
\right\rangle
\eea
where $\kappa_{n}= \frac{\pi^{2}T\,n}{6\Gamma m}$ is the thermal conductivity in the 
normal state.  Then in the superclean limit Eq.(56) reduces to
\bea
\kappa_{\parallel}/\kappa_{n} &=& \frac{\pi\Gamma}{2\Delta}\left\langle \frac{\theta(x^{2}-f^{2})}
{x C_{0}}(x^{2}-f^{2})^{1/2}\right\rangle \\
&=& \frac{\pi\Gamma}{4\Delta C_{0}}<x> = \frac{\pi}{4}<x>^{2} \\
&=& \frac{1}{\pi^{3}}\frac{v^{2}(eH)}{\Delta^{2}}
\eea
In the last step we assumed ${\bf H} \parallel {\bf c}$.  The H-linear thermal conductivity
was first observed in Sr$_{2}$RuO$_{4}$ at $T =0.3 K$ [\onlinecite{izawa-2}], which indicates nodal superconductivity
with $f=e^{\pm i\phi}\cos(\chi)$ in Sr$_{2}$RuO$_{4}$ and that the system is in the superclean limit.  More
recently the H-linear thermal conductivity was observed in 
PrOs$_{4}$Sb$_{12}$ [\onlinecite{maki-2}] for $T \leq 0.3 K$.

Now in the presence of a magnetic field in the a-b plane, we must first generalize 
Eq. 57 as [\onlinecite{won-4}]
\bea
\kappa_{xx}/\kappa_{n} &=& \frac{\pi}{4}<x><(1+\cos(2\phi^{'}))x> \\
\kappa_{yy}/\kappa_{n} &=& \frac{\pi}{4}<x><(1-\cos(2\phi^{'}))x>
\eea and
\bea
\kappa_{xy}/\kappa_{n} &=& \frac{\pi}{4}<x><\sin(2\phi^{'})x>
\eea
where the angle $\phi^{'}$ refers to the direction of the quasiparticle wave vector within the
a-b plane.  Then for $f=\cos(2\phi)$ Eq.(58) gives
\bea
 \kappa_{\parallel}/\kappa_{n} ( \equiv \kappa_{xx}/\kappa_{n}&=& \kappa_{yy}/\kappa_{n}) = \frac{{\tilde{v}}^{2}eH}{\pi^{3}\Delta^{2}}
(0.955+0.0285\cos(4\phi))^{2}
\eea
and
\bea
\kappa_{xy}/\kappa_{n}&=& -\frac{{\tilde{v}}^{2}eH}{\pi^{3}\Delta^{2}}(0.265\sin(2\phi)
(0.955++0.0285\cos(4\phi))
\eea
Here $\tilde{v}= \sqrt{v v_{c}}$.  Therefore $\kappa_{\parallel}$ should exhibit the fourfold term
with a magnitude variation of the order of $\sim~ 6 \%$.

In early experiments [\onlinecite{yu,aubin,106}] a fourfold term of comparative magnitude was found in YBCO but
of opposite sign.  Similarly Oca\~{n}a and Esquinazi [\onlinecite{ocana}] found the Hall thermal
conductivity $\sim \sin(2\phi)$, but of opposite sign.  These problems were clarified in [\onlinecite{won-5}].
All experiments on YBCO were performed at $T \gg <|{\bf v}\cdot{\bf q}|>$, whereas the present
theory applies only for $T \ll <|{\bf v}\cdot{\bf q}|>$.

Also, thermal conductivity measurements in CeCoIn$_{5}$ [\onlinecite{izawa-3}] and 
$\kappa$-(ET)$_{2}$Cu(NCS)$_{2}$ [\onlinecite{izawa-4}] 
have seen evidence for a fourfold term.  If we assume that these experiments are done in 
the region $T \ll <|{\bf v}\cdot{\bf q}|>$, we have to interpret the data as $d_{xy}$-wave
superconductivity and $d_{x^{2}-y^{2}}$ symmetry for CeCoIn$_{5}$ and $\kappa$-(ET)$_{2}$Cu(NCS)$_{2}$
respectively.  The former identification appears to be consistent with the magnetospecific
heat data of CeCoIn$_{5}$ reported in [\onlinecite{107}].  For $f=e^{\pm i\phi}\cos \chi$, we obtain 
[\onlinecite{dahm}]
\bea
\kappa_{xx}/\kappa_{n}&=& \frac{\tilde{v}^{2}(eH)}{\Delta^{2}}\times 1.479(1-0.0416\cos(2\phi))
\eea
The magnitude of the twofold term seen in [\onlinecite{izawa-1}] is consistent with Eq.(66).

\subsection{Clean limit}

In the clean limit Eq.(57) is transformed as
\bea
\kappa_{\parallel}/\kappa_{0} &=& 1 +\frac{<x^{2}>}{3 c_{0}^{2}} \\
& \simeq & 1 + \frac{1}{6\pi^{2}}\ln\left(2\sqrt{\frac{2\Delta}{\pi\Gamma}}\right)\frac{v^{2}(eH)}{\Gamma \Delta}
\ln(\frac{\Delta}{v\sqrt{eH}})
\eea
where $\kappa_{0}=\kappa_{\parallel}(H \rightarrow 0)$ and we have assumed ${\bf H} \parallel {\bf c}$.
In the clean limit the field dependence is given by $\kappa_{\parallel} \sim H \ln(H_{0}/H)$.
In a magnetic field within the a-b plane we obtain
\bea
\kappa_{xx}/\kappa_{0} &=& 1 + \frac{1}{6\pi^{2}}\ln\left(2\sqrt{\frac{2\Delta}{\pi\Gamma}}\right)
\frac{{\tilde{v}}^{2}(eH)}{\Gamma\Delta} [\ln(\frac{\Delta}{{\tilde{v}}\sqrt{eH}})-0.072+0.041\cos(4\phi)]
\eea
and
\bea
\kappa_{xy}/\kappa_{0} &=& \frac{1}{3c_{0}^{2}}<\sin(2\phi)x^{2}> \\
&=&-\frac{(\tilde{v})^{2}eH}{3\pi^{2}\Gamma\Delta}\ln(2\sqrt{\frac{2\Delta}{\pi \Gamma}})\sin(2\phi)
[\ln(\frac{\Delta}{\tilde{v}\sqrt{eH}})-0.42]
\eea
where we have assumed $f=\cos(2\phi)$.

The b-axis thermal conductivity of $\kappa$-(ET)$_{2}$Cu(NCS)$_{2}$ determined by Izawa et al
[\onlinecite{izawa-3}] exhibits not only the four-fold term predicted in Eq. 65, but also a two-fold term
that was not predicted.  We can interpret the two-fold term as being due to the admixture of an
s-wave component.  The experimental data is most naturally interpreted as $\Delta({\bf k})
\sim \cos(2\phi) - 0.067$ [\onlinecite{108}].  Finally for $T \gg <|{\bf v}\cdot{\bf q}|>$ we obtain
[\onlinecite{won-5}]
\bea
\kappa_{xx}/\kappa_{n} = \frac{7\pi^{2}}{10}(\frac{T}{\Delta})^{2}(1+(\frac{2}{\pi})^{2}\ln^{2}
(\frac{2\Delta}{1.76 T})) - \nonumber \\
\frac{1}{(2\pi)^{2}}\ln(\frac{2\Delta}{1.76 T})\frac{\tilde{v}^{2}(eH)}
{\Delta^{2}}\left[\ln(\frac{4\Delta}{\tilde{v}\sqrt{eH}}-\frac{1}{16}(1-\cos(4\phi))\right] \\
\kappa_{xy}/\kappa_{n} = \frac{1}{2\pi^{2}}\sin(2\phi)\frac{\tilde{v}^{2}(eH)}{\Delta^{2}}
\ln(\frac{2\Delta}{1.76T})\ln(\frac{4\Delta}{\tilde{v}\sqrt{eH}})
\eea
It appears that both Eq.(67) and (68) describe consistently the angle-dependent
magnetothermal conductivity observed in YBCO [\onlinecite{ocana,106}].

{\bf Exercises 5.}\\
5.1 Indicate how to derive Eq.(58) from Eq.(57).\\
5.2 Consider $\kappa_{zz}$ for f's with horizontal nodes (i.e. $f=\sin\chi$, $\cos\chi$, and
$\cos 2\chi$) when the magnetic field is rotated within the z-x plane.

\section{Superconductivity with Point Nodes}

\subsection{Borocarbides YNi$_2$B$_2$C and LuNi$_2$B$_2$C}

This class of superconductivity was discovered in 1994 [\onlinecite{109}]. Their relatively high transition temperatures
(T$_c$ = 15.5 K and 16.5 K for YNi$_2$B$_2$C and LuNi$_2$B$_2$C, respectively)
and their interplay between magnetism and superconductivity 
are of great interest. In the following we will focus on the two superconducting 
borocarbides YNi$_2$B$_2$C and LuNi$_2$B$_2$C, which have no
indication of magnetism.

The presence of a substantial s-wave component in the
superconducting order parameter $\Delta({\bf k})$ in
YNi$_2$B$_2$C was established by substituting Ni by a small
amount of Pt.  An opening of the energy gap was observed by specific
heat measurements [\onlinecite{110}].  On the other hand, this superconductivity exhibits 
a number of peculiarities unexpected
for s-wave superconductors.  For example the $\sqrt{H}$ dependence 
of the specific heat in the vortex state [\onlinecite{111,112}] indicates a nodal superconductor.  Furthermore
the presence of de-Haas-van-Alphen (dHvA) oscillation in the vortex
state in LuNi$_2$B$_2$C down to $H=0.2 H_{c2}$ suggests again nodal
superconductivity.[\onlinecite{113,114}]  Also
$T_{1}^{-1}$ in NMR exhibits a $T^{3}$ power law consistent with
nodal superconductivity [\onlinecite{115}].
In addition $\Delta({\bf k})$ exhibits a
fourfold symmetry as seen from the angular dependence of the
upper critical field $H_{c2}$ when the magnetic field is 
rotated within the a-b plane [\onlinecite{116,117}]. Also, unlike
the superconductors discussed in previous sections, here we are dealing with
a superconductor with a 3D Fermi surface.

Then we postulate [\onlinecite{maki-6}]
\bea
\Delta({\bf k}) = (\Delta/2)(1-\sin^{4}(\theta)\cos(4\phi))
\eea
where $\theta$ and $\phi$ are polar coordinates describing
${\bf k}$, the quasiparticle wave vector.  This order parameter belongs to the hybrid
representation of s+g-wave superconductivity.  The precise matching
of the s and g-wave components is necessary in order to find the nodal
excitations as observed experimentally.  Within a model Hamiltonian
the stability of such a precise matching is considered in [\onlinecite{119}].  The corresponding
$\Delta({\bf k})$ is shown in Fig. 13.  The quasiparticle density of states is
\begin{figure}[ht]
\includegraphics[width=6cm]{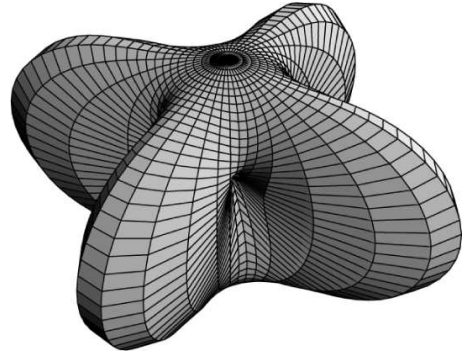}
\caption{Order parameter for the borocarbides YNi$_2$B$_2$C and LuNi$_2$B$_2$C}
\end{figure}
given by
\bea 
    G(E) = |x|\int_{0}^{2}dy\,F(y)Re\frac{1}{(x^{2}-y^{2})^{1/2}}
\eea
where $x=E/\Delta$ and 
\bea    
 F(y)= \frac{1}{2}\int_{0}^{u_{0}}\,\frac{dz}{((1-z^{2})^{4}-(1-u_0^{2})^{4})}
\eea 
with $u_0= \sqrt{( 1-(|1-y|)^{1/2})}$. We note that $F(2-y)=
F(y)$. The DOS is shown in Fig. 14.  For $|E|/\Delta = x \ll 1$ the quasiparticle density of
\begin{figure}[ht]
\includegraphics[width=8.5cm]{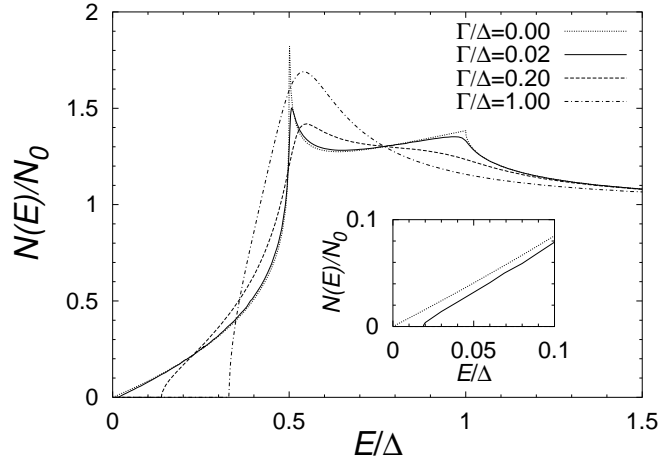}
\caption{ Quasi-particle density of states for the borocarbides YNi$_2$B$_2$C and LuNi$_2$B$_2$C}
\end{figure}
states can be approximated by
\bea
g(E) = \frac{\pi |E|}{8 \Delta}(1+(9/(8\pi))(|E|/\Delta)+\ldots)
\eea        
which gives
\bea
C_{s}/\gamma_{N}T & \simeq & \frac{27 \zeta(3)}{8 \pi}(T/\Delta)
\eea
Unlike the point nodes discussed in Ref. [\onlinecite{sigrist}], the point nodes here are quadratic
which gives $N(E) \sim |E|$ and $C_{s} \sim T^{2}$.
In the presence of a magnetic field the quasiparticle density of
states is given by 
\bea
g(0,{\bf H}) = \frac{\pi}{8\Delta}
<|{\bf v}\cdot {\bf q}|> = \frac{\tilde{v}(eH)^{1/2}I(\theta,\phi)}{2\Delta}
\eea
where $\tilde{v} = \sqrt{v_{a}v_{c}}$ and
\bea
  I(\theta,\phi)= \frac{1}{2}((1-\sin^{2}\theta \sin^{2}\phi)^{1/2} +(1-\sin^{2}\theta \cos^{2}\phi)^{1/2})
\eea
Here $\theta$ and $\phi$ are the direction of the magnetic field
where the polar axis is taken parallel to the c axis. We show
in Fig. 15 $I(\theta,\phi)$.  In particular for $\theta=\pi/2$ 
\begin{figure}[h]
\includegraphics[width=8.5cm]{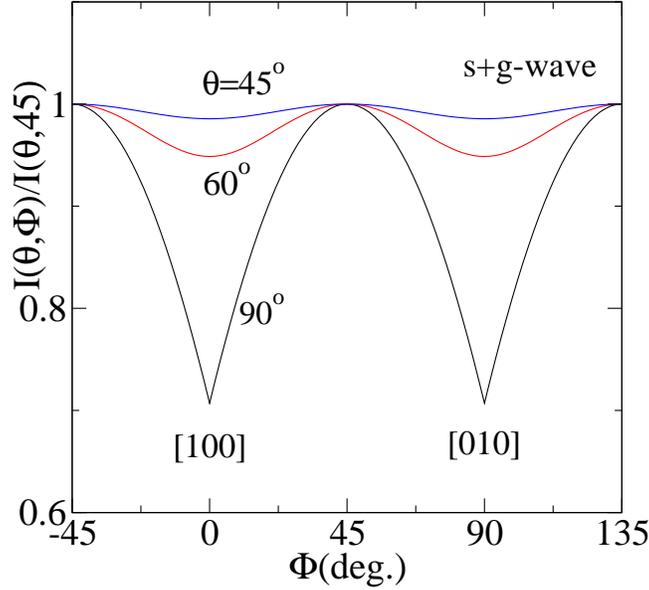} 
\caption{The angular function $I(\theta,\phi)$}
\end{figure}
the DOS exhibits cusps at $\phi = 0,\pi/2,\pi,3\pi/2$.  These cusps appear
also in the specific heat, since 
\bea
C_{s}/\gamma_{N}T &=& g(0,H)
\eea
for $T \ll \Delta$.  Indeed, such cusps have been seen by Park et al [\onlinecite{120}.]  
Unfortunately, however, their data are limited to $\theta=\pi/2$ and cannot
explore the whole extent of I($\theta,\phi$).

Due to the presence of an s-wave component, the effect of impurity scattering is
very different from the usual nodal superconductors discussed in {\bf III} and {\bf IV}.[\onlinecite{121}]  First
of all there is no resonant scattering.  So the Born approximation suffices.  
Also, unlike in other nodal superconductors the energy gap opens up immediately in the presence
of impurities.  The energy gap $\omega_{g}$ is given in a good approximation by
$\omega_{g}=\Gamma(1+\frac{\Gamma}{\Delta})^{-1}$.  Therefore the specific heat and the thermal
conductivity decrease as $T \rightarrow 0$ like $C_{s}/T \sim \kappa/T \sim (\omega_{g}/T)^{2}
e^{-\omega_{g}/T}$.  Hence, there is no universal heat conduction.

The gap equation is solved in the presence of impurities.  We find
\bea
 -\ln(T_c/T_{c0})=\frac{0.203}{1.203}[\Psi(\frac{1}{2}+\frac{\Gamma}{2\pi T_c})-\Psi(\frac{1}{2})]
\eea
which is compared with Eq.(37) for usual nodal superconductors.  We 
show in Fig. 16 $T_{c}/T_{c0}, \Delta(0,\Gamma)/\Delta_{00}$ versus $\Gamma/\Delta_{00}$.  Both 
$T_{c}/T_{c0}$ and, $\Delta(0,\Gamma)/\Delta_{00}$ decrease much more slowly as $\Gamma/\Delta_{00}$
increases.
\begin{figure}[h]
\includegraphics[width=10cm]{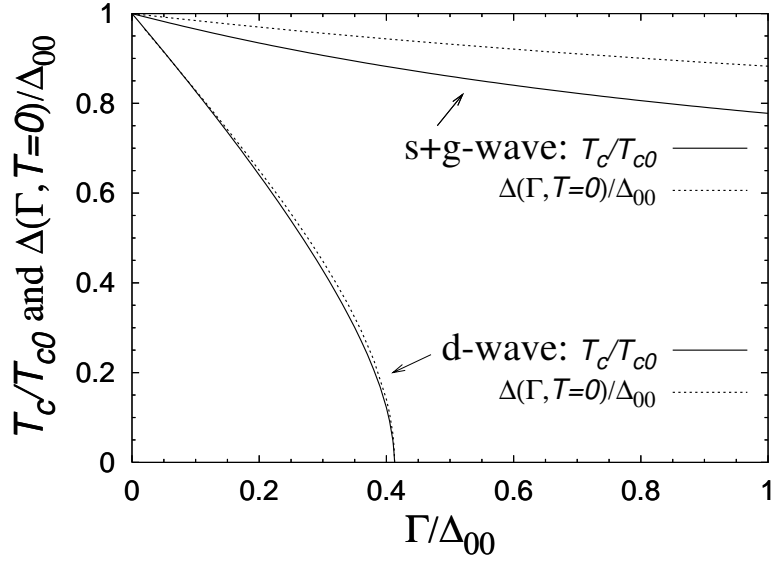}
\caption{T$_c$ and $\Delta(0,\Gamma)$ for borocarbides with impurities}
\end{figure}

Finally the thermal conductivity in the vortex state is given by [\onlinecite{122}]
\bea
\kappa_{zz}/\kappa_n & = & \frac{x}{2\ln(2/x)}[(1-y^{2})^{3/2}-
\frac{3}{2}y(\arccos y-y(1-y^{2})^{1/2})]\theta(1-y) \\
& \simeq & \frac{x}{2\ln(2/x)} 
\eea
where $x = \frac{\tilde{v}(eH)^{1/2} I(\theta,\phi)}{\Delta},  y= \frac{\Gamma}{\Delta x}$.
We show in Fig. 17 data by Izawa at al [\onlinecite{izawa-4}], where the magnetic field
is rotated conically around the c-axis.  The clear cusps for $\theta=\pi/2$ disappear
as $\theta$ is decreased.  This is the clear sign of point nodes at ${\bf k}= (1 0 0), (0 1 0),
(-1 0 0)$ and $(0 -1 0)$.
\begin{figure}
\includegraphics[width=8.5cm]{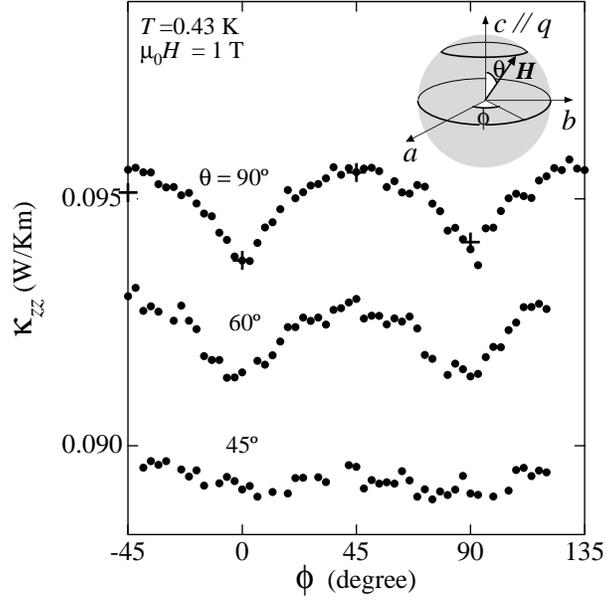}
\caption{Thermal conductivity in vortex state in YNi$_2$B$_2$}
\end{figure}
The effect of Pt-substitution of Ni in Y(Ni$_{1-x}$Pt$_x$)$_2$B$_2$C
with x=0.05 is studied by Kamata et al [\onlinecite{123}].  As seen from Fig. 19
\begin{figure}[h]
\includegraphics[width=8.5cm]{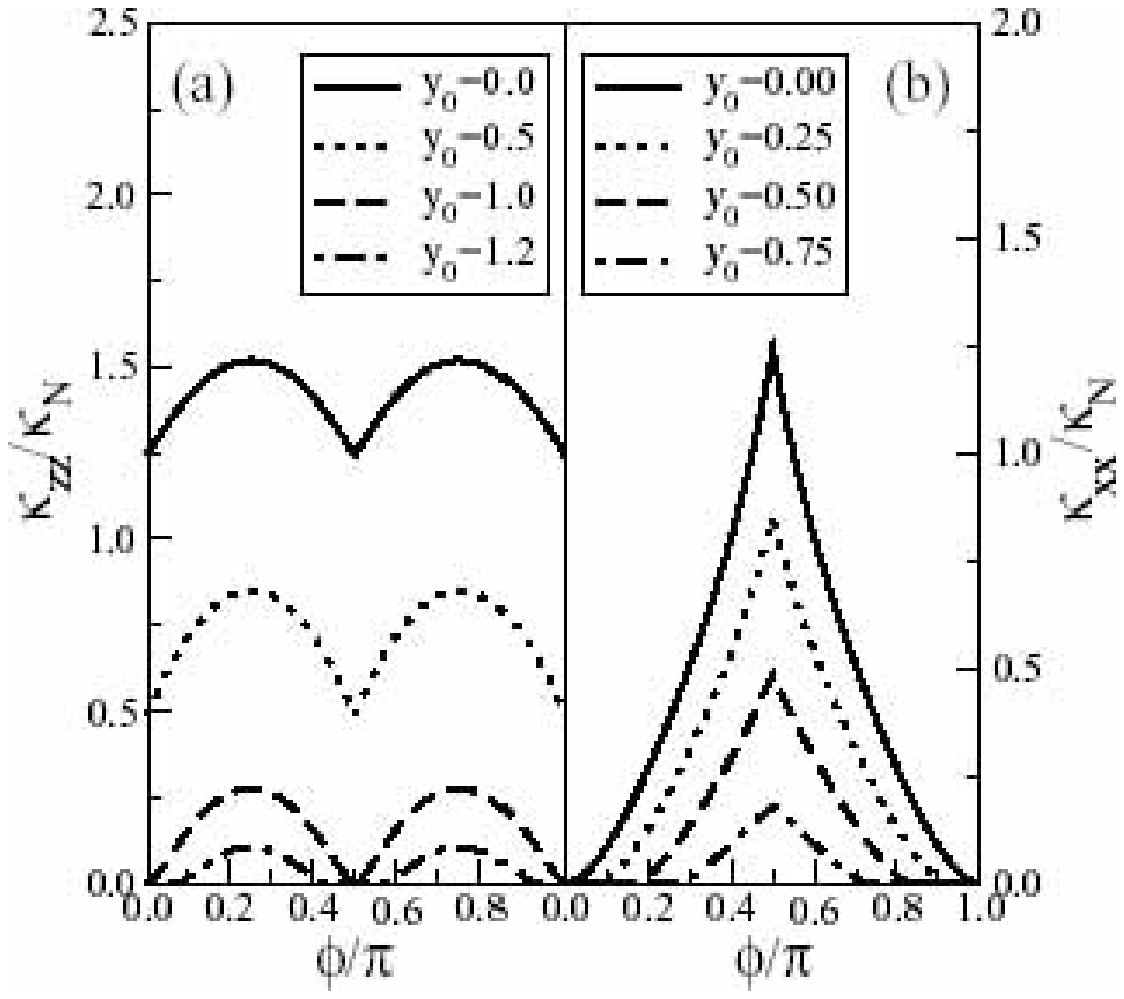}
\caption{Effect of impurities on thermal conductivity - theoretical}
\includegraphics[width=8.5cm]{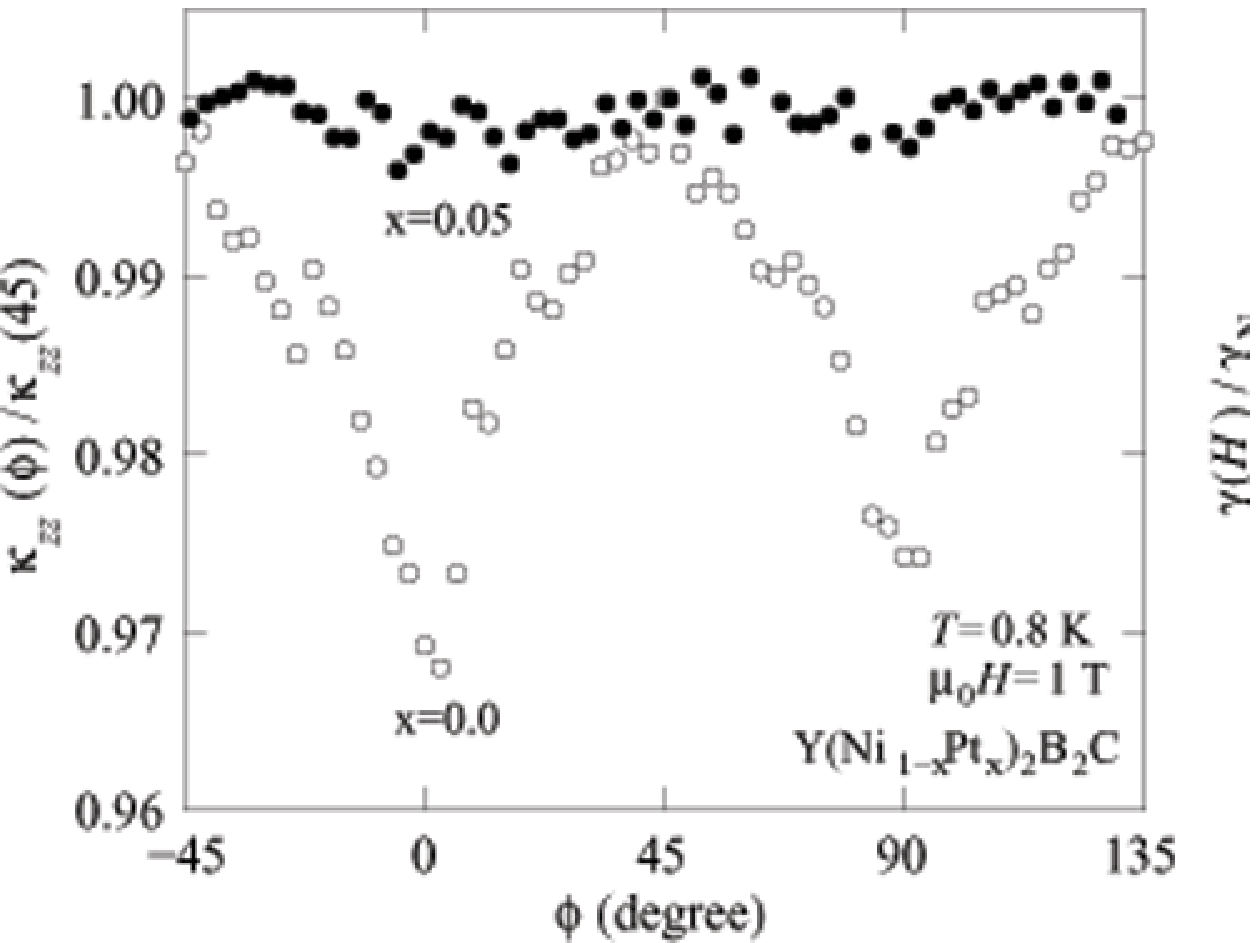}
\caption{Effect of impurities on thermal conductivity - experimental}
\end{figure}the angular dependence in $\kappa_{zz}$ disappeared completely for 5\% substitution
of Pt.  From Eq.(78) and $T_{c}=13.1 K$ of
the 5 percent Pt substituted system, we can deduce $\Gamma= 23.8 K$.
This $\Gamma$ is much larger than T$_{c0}$. Clearly parallel experiments with Pt-substitution
less than 1\% are highly desirable.

The present $\Delta({\bf k})$ describes very well the ultrasonic attenuation data of YNi$_{2}$B$_{2}$C
[\onlinecite{124,125}].

\subsection{Skutterudite PrOs$_{4}$Sb$_{12}$}

The skutterudite PrOs$_4$Sb$_{12}$ is a heavy-fermion superconductor
with transition temperature T$_{c} \sim 1.8 K$ [\onlinecite{126,127,128}].  The angle-dependent
magnetothermal conductivity data of this material inidcates an interesting
multiphase structure characterized by the gap functions $\Delta_{A}({\bf k})$ and
$\Delta_{B}({\bf k})$ with point nodes.[\onlinecite{izawa-5,maki-2,129}] More recently, there has been mounting
evidence for triplet superconductivity in this compound.  First, from $\mu$-SR measurement
Aoki et al [\onlinecite{130}] discovered a remnant magnetization in the B phase of this compound
indicating triplet pairing.  Also more recent thermal conductivity data by Izawa et al
is not consistent with the singlet model [\onlinecite{129}], but is consistent 
with p+h-wave superconductors 
[\onlinecite{maki-2}].  Finally, Tou et al [\onlinecite{131}] reported NMR data of PrOs$_{4}$Sb$_{12}$ of which the Knight
shift suggests the triplet pairing.

The triplet pairing together with the position of point nodes give almost uniquely [\onlinecite{maki-2}]
\bea
{\bf \Delta_{A}}({\bf k}) &=& \frac{3}{2}{\bf d}\Delta e^{\pm i \phi_{i}}(1-{\bf k_{1}}^{4}-
{\bf k_{2}}^{4} - {\bf k_{3}}^{4}) \\
{\bf \Delta_{B}}({\bf k}) &=& {\bf d}\Delta e^{\pm i \phi_{3}}(1-{\bf k_{3}}^{4})
\eea
where $e^{\pm i \phi_{1}} = (k_{2} \pm i k_{3})/\sqrt{(k_{2}^{2}+k_{3}^{2})}$, etc.
Here $\phi_{i}$ in Eq.(85) is one of $\phi_{1}, \phi_{2}$ and $\phi_{3}$.  Also
we have chosen the nodal direction of ${\bf \Delta_{B}}$ parallel to (0 0 1) in Eq.(86).
Also $|{\bf \Delta_{A}}({\bf k})|$ has the cubic symmetry consistent with the crystal symmetry
of PrOs$_4$Sb$_{12}$, while $|{\bf \Delta_{B}}({\bf k}|$ has the axial symmetry.  We show in Fig. 20
$|{\bf \Delta_{A}}({\bf k}|$ and $|{\bf \Delta_{B}}({\bf k})|$.
\begin{figure}[ht]
\includegraphics[width=6cm]{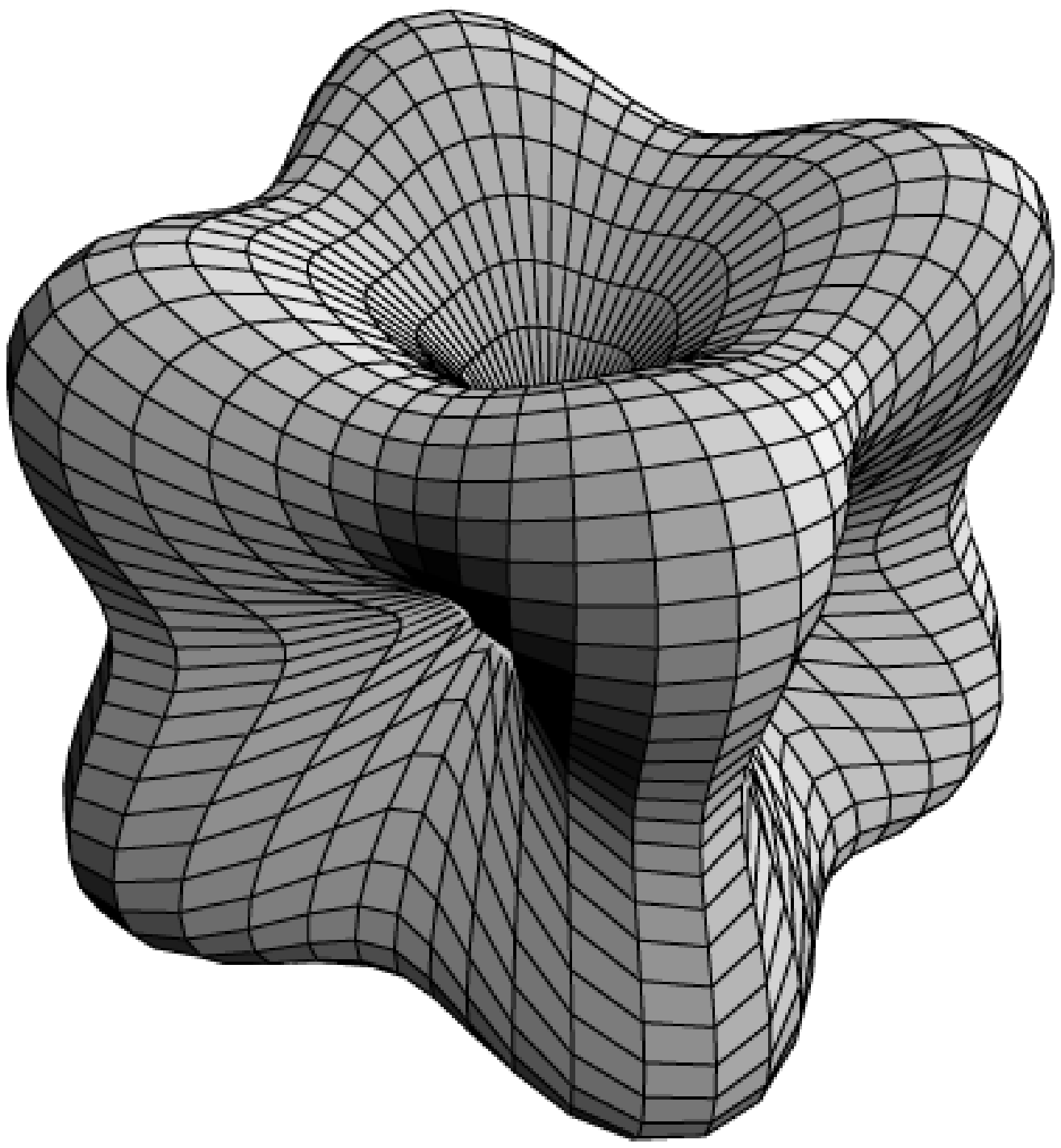}
\includegraphics[width=6cm]{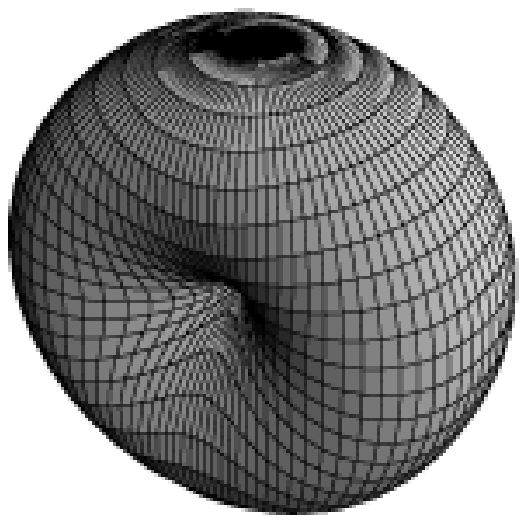}
\caption{Proposed A-phase (left) and B-phase order parameters for PrOs$_{4}$Sb$_{12}$}
\end{figure}
Very recently the magnetic penetration depth of the B-phase in PrOs$_4$Sb$_{12}$ has been reported
[\onlinecite{132}].  They applied a magnetic field parallel to each of the 3 crystal axes and 
determined the magnetic penetration depth and found isotropic superfluid density.  At first sight
this is clearly in contradiction to our ${\bf \Delta_{B}}({\bf k})$ in Eq. (86).  However,
this is understood, if we assume that the stationary magnetic field controls the 
symmetry axis of ${\bf \Delta_{B}}({\bf k})$ [\onlinecite{133}].  Indeed the ground state energy is favorable
when the nodes are aligned parallel to ${\bf H}$.  Also the $T^{2}$ dependence of the
superfluid density for $T \ll \Delta$ follows from this assumption.  As in all the triplet
superconductors, our superconducting order parameter breaks the chiral symmetry.  So both order
parameters are six-fold degenerate.

The quasiparticle DOS of these states is given by
\bea
g(E) &=& |x| Re \left<\frac{1}{\sqrt{x^{2}-|f|^{2}}} \right >
\eea
for $x=\frac{|E|}{\Delta}$.  This is evaluated numerically and shown in Fig. 21.
\begin{figure}[h]
\includegraphics[width=8.5cm]{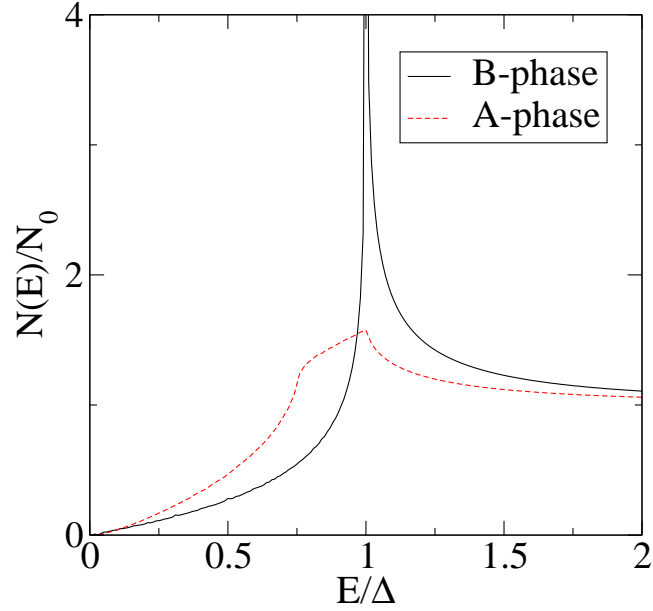}
\caption{Quasiparticle density of states for p+h-wave superconductor}
\end{figure} For $|x| \ll 1$ we obtain
\bea
g(E) & \simeq & \pi |x|/4 \,\,\,\,\mathrm{A-phase}  \\
& \simeq & \pi |x|/8 \,\,\,\,\mathrm{B-phase}
\eea
In the B-phase there is a logarithmic singularity at $E=\Delta$ as in d-wave
superconductors, whereas in the A-phase the singularity is split into 2 cusps at 
$E/\Delta = 3/4$ and $1$.  In the vortex state the quasiparticle DOS is given by
\bea
g_{A}({\bf H}) & = & \frac{1}{2}\frac{v\sqrt{eH}}{\Delta} I_{A}(\theta,\phi) \,\,\,\,\mathrm{A-phase} \\
&=& \frac{1}{4}\frac{v\sqrt{eH}}{\Delta} I_{B}(\theta,\phi) \,\,\,\,\mathrm{B-phase}
\eea
where
\bea
I_{A}(\theta,\phi) &=& \sin\theta + (1-\cos^{2}\theta \sin^{2} \phi)^{1/2} + (1-\cos^{2}\theta \cos^{2} \phi)^{1/2}  \\
I_{B}(\theta,\phi) &=& \sin \theta.
\eea
In the B-phase we have assumed that the nodes are parallel to (0 0 1).
Then the low-temperature specific heat, etc. are given by
\bea
C_{S}/\gamma_{N}T &=& g({\bf H}), \,\,\,\, \chi_{S}/\chi_{N}= g({\bf H}) \\
\rho_{S}({\bf H})/\rho_{S}(0) &=& 1 - g({\bf H})\,\,\,\, (\mathrm{A-phase}) \\
\rho_{S\parallel}({\bf H})/\rho_{S\parallel}(0) &=& 1 - 3 g({\bf H})\,\,\,\, (\mathrm{B-phase})
\eea
The last expression means the superfluid density measured parallel to the nodal directions.

In p+h-wave superconductors the impurity scattering is handled similarly to other nodal
superconductors.  For example, the superconductivity in the A-phase exhibits the 
universal heat conduction with $\kappa_{00}/T = \pi^{2}nE_{F}/12\Delta(0)$.  On the other hand
in the B-phase the heat current has to be parallel to the nodal direction.  Then we will
have $\kappa_{00}/T = \pi^{2}nE_{F}/8\Delta(0)$.  When the heat current is perpendicular to the
nodal direction in the B-phase, the thermal conductivity vanishes like $T^{2}$.

Now let us consider $\kappa_{zz}$ in the vortex state when ${\bf H}$ is rotated in the
z-x plane with $\theta$ the angle {\bf H} makes from the z axis.  Then in the superclean limit
($(\Gamma \Delta)^{1/2} \ll v\sqrt{eH}$) we obtain
\bea
\frac{\kappa_{zz}}{\kappa_{n}} &=& \frac{v^{2}eH}{8\Delta^{2}} 
\sin^{2}\theta \,\,\,\,\mathrm{A-phase} \\
&=&\frac{3v^{2}eH}{64\Delta^{2}} \sin^{2}\theta \,\,\,\,\mathrm{B-phase}
\eea
where in the B-phase we assumed that the nodes are parallel to (0 0 1).
Similarly in the clean limit ($(\Gamma\Delta)^{1/2} \gg v\sqrt{eH}$) we obtain
\bea
\kappa_{zz}/\kappa_{0} &=& 1 + \frac{3v^{2}eH}{40\Gamma\Delta}\ln(\sqrt{\frac{2\Delta}{\Gamma}})
\sin^{2}\theta \ln(\frac{\Delta}{v\sqrt{eH}\sin \theta})\,\,\,\, (\mathrm{A-phase}) \\
&=& 1 + \frac{v^{2}eH}{12\Gamma\Delta}\ln(\sqrt{\frac{2\Delta}{\Gamma}})
\sin^{2}\theta \ln(\frac{\Delta}{v\sqrt{eH}\sin \theta})\,\,\,\, (\mathrm{B-phase})
\eea
where $\kappa_{0} = \kappa_{zz}(H=0)$.  In Fig. 22 we show the experimental data at
T = 0.35 K together with the theoretical fit from Eq.(90) and (91).  As is readily seen
\begin{figure}[h]
\includegraphics[width=10.5cm]{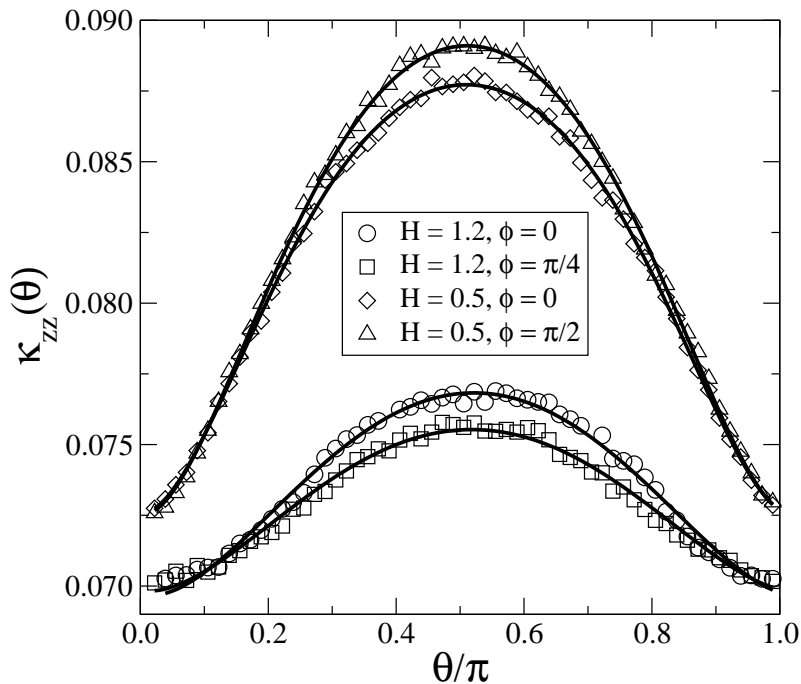}
\caption{Thermal conductivity of PrOs$_{4}$Sb$_{12}$ in the vortex state}
\end{figure}
the fits are excellent with $\Delta/v\sqrt{eH}$ = 5 and 3 for data at H = 0.5 T (B phase)
and H= 1.2 T (A-phase) respectively.  Then making use of the weak-coupling theory gaps
$\Delta_{A}(0)=4.2 K$ and $\Delta_{B}(0)=3.5 K$ for the A-phase and B-phase respectively,
we can deduce v = $0.96 \times 10^{7}$ cm/sec and $\Gamma \simeq 0.1 K$.  These values are
very reasonable.  From the de Haas-van Alphen measurement v is estimated to be $0.7 
\times 10^{7}$ cm/sec ($\alpha$ band), $0.66 \times 10^{7}$ cm/sec ($\beta$ band), and 0.23
$\times 10^{7}$ cm/sec ($\gamma$ band) [\onlinecite{134}].  Perhaps the triplet superconductivity in PrOs$_4$Sb$_{12}$
is not so surprising.  The superconductivity in Sr$_{2}$RuO$_{4}$, UPt$_{3}$, UBe$_{13}$, 
URu$_{2}$Si$_{2}$ and UNi$_{2}$Al$_{3}$ appears to be triplet [\onlinecite{135}].  More surprising is the high
degeneracy of the ground state due to the chiral symmetry-breaking.  Both the A-phase and the B-phase
have six-fold degeneracy.  This means that a variety of topological defects in both the A-phase
and the B-phase are likely, which deserve further study.  The multi-phase structure is rare but
at least we have seen it in superfluid ${}^{3}$He and in UPt$_{3}$.

On the other hand the exact location of the A-B phase boundary is still controversial.  A recent
thermodynamical study [\onlinecite{136}] indicates that the A-B phase boundary is rather parallel to the
upper critical field H$_{c2}$ of the A phase.  Somewhat surprisingly the less symmetric B phase
is realized for the low-field region.  Is this related to the antiferromagnetic quadrupolar
(AFQ) state which appears above $H \simeq 4 T$ [\onlinecite{137}]? It looks like both the
superconductivity in YNi$_{2}$B$_{2}$C and in PrOs$_{4}$Sb$_{12}$ have great futures.

\subsection{Summary on $\Delta({\bf k})$}

a). Quasi-2D systems

1. d-wave superconductors ($f = \cos(2\phi)$; hole doped high-T$_{c}$ cuprates (YBCO, 
LSCO, Bi2212, Tl2201), electron-doped cuprates (NCCO, PCCO), CeCoIn$_{5}$ and 
$\kappa-(ET)_{2}Cu(NCS)_{2}$.  As to the gap symmetry of CeCoIn$_{5}$ there is controversy
between d$_{x^{2}-y^{2}}$-wave versus d$_{xy}$-wave.  

2. Sr$_{2}$RuO$_{4}$.  Early experiments established triplet pairing and chiral
symmetry breaking [\onlinecite{138,139,140}].  When the high quality single crystals became available,
both the specific heat data [\onlinecite{nishizaki}] and the magnetic 
penetration data [\onlinecite{141}] exhibited
characteristics of nodal superconductors.  Therefore the p-wave superconductivity proposed
in [\onlinecite{142}] is clearly out.  In order to save this situation
a multigap model was proposed by Zhitomirsky and Rice [\onlinecite{143}].  However, our analysis of
the optical conductivity [\onlinecite{145}] indicates that there is little room for p-wave
superconductivity.  Therefore the important question is: where are the line nodes in $\Delta({\bf k})$.

Clearly the magnetothermal conductivity data [\onlinecite{izawa-1}] and the ultrasonic attenuation data
[\onlinecite{144}] support horizontal nodes as in $f = e^{\pm i \phi}\cos \chi$.  Therefore, except
for the specific heat data by Deguchi et al [\onlinecite{deguchi}] we mentioned earlier all available data
are consistent with the 2D f-wave model.  Also we have 
proposed that the optical conductivity [\onlinecite{145}],
the Raman spectra [\onlinecite{146}], and the supercurrent 
experiment [\onlinecite{147,148}] below T =0.1 K will provide the
definitive test of $\Delta({\bf k})$ in Sr$_{2}$RuO$_{4}$.  The review paper by Mackenzie and Maeno
[\onlinecite{149}] is excellent, but does not cite the significant evidence
for nodal superconductivity in Sr$_{2}$RuO$_{4}$, if we concentrate on the data from high quality
single crystals.  On the other hand, we agree that definitive experiments below $T= 100$ mK are
highly desirable.

3. UPd$_{2}$Al$_{3}$  As briefly described in {\bf IV}, 
we have proposed f = $\cos(2\chi)$.[\onlinecite{won-new}]

4. We are awaiting the magnetothermal conductivity data of UNi$_{2}$Al$_{3}$, URu$_{2}$Si$_{2}$ and
CeCu$_{2}$Si$_{2}$.

5) 3D systems

a).  YNi$_{2}$B$_{2}$C, LuNi$_{2}$B$_{2}$C.  $\Delta({\bf k})= (\Delta/2)(1-\sin^{4}\theta \cos(4\phi))$

b).  PrOs$_{4}$Sb$_{12}$, p+h-wave superconductivity.

\section{D-wave Density Wave}

In the preceding lectures we have reviewed a variety of nodal superconductors which
have appeared since 1979.  For some of them we have succeeded in identifying the gap
symmetry.  All these analyses are based on the BCS theory of nodal superconductors.
Also the semiclassical approximation played an important part in elucidating the
quasiparticle properties in the vortex state.

Therefore it is very natural to contemplate a parallel development in charge density wave (CDW)
and spin density wave [\onlinecite{ishizuro,150}] systems   The unconventional density wave (UDW) was speculated as a 
possible electronic ground state in excitonic insulators in 1968 [\onlinecite{151}].  However, the recent surge
of interest in UDW is in part due to the proposal that the pseudogap phase in high-T$_{c}$ cuprate
superconductors is a d-wave density wave (d-DW).[\onlinecite{cappelutti,benfatto,chakravarty,dora-1}]  
It is important to point out that
two kinds of d-DW have been considered.  The first one is commensurate and individual square lattice
enclosing a circulating current [\onlinecite{cappelutti,chakravarty}] analogous to the flux phase 
[\onlinecite{affleck}].  Therefore the ground state
has Z$_{2}$ symmetry and ``visons'' as topological defects. [\onlinecite{senthil}] On the other hand we consider
a d-DW, which is in general incommensurate. [\onlinecite{dora-1}] Therefore as in a superconductor the ground state
has U(1) symmetry associated with the axial gauge transformation
\bea
c_{k}^{\dagger} \rightarrow e^{i\phi/2}c_{k}^{\dagger},\,\,\,\, c_{k+Q}^{\dagger} \rightarrow 
e^{-i\phi/2}c_{k+Q}^{\dagger}
\eea
The d-DW can have ``phase vortices'' similar to the ones in conventional CDW [\onlinecite{ong}].

Here are a few characteristics of unconventional density waves (UDW) [\onlinecite{152}].  First of all,
the transition from the normal state to UDW is a metal-to-metal transition.  Though the quasiparticle
density decreases due to the opening of a partial energy gap,  UDW are conductors down to T = 0 K.  In
high-T$_{c}$ cuprates the resistivity $\sim T^{2}$ in the normal state changes to the resistivity
$\sim T$ for example.  Also, we shall see that the quasiparticles in UDW are a standard Fermi liquid.
Therefore this anomalous resistivity behavior does not imply the existence of a ``non-Fermi liquid''.
Since $\langle\Delta({\bf k})\rangle $=0, there will be no x-ray or neutron signal for charge density or
spin density.  Therefore UDW is sometimes called a condensate with a hidden order parameter
[\onlinecite{chakravarty}].

The quasiparticle Green function in UDW is given by
\bea
\mathcal{G}^{-1}(k,\omega) = \omega - \eta(k)- \xi^{'}(k)\rho_{3}-\Delta(k)\rho_{1}
\eea
where $\xi^{'}(k)= \frac{1}{2}(\xi(k)-\xi(k-Q)), \eta(k)=\frac{1}{2}(\xi(k)+\xi(k+q))$
and ${\bf Q}$ is the nesting vector.  Also the $\rho 's$ are the Pauli matrices operating on
the spinor space made of $|c_{k}^{\dagger}\rangle$ and $|c_{k-Q}^{\dagger}\rangle$.  Here we consider
only UCDW for simplicity and spin indices are dropped.  In the following we drop the prime
for $\xi^{'}(k)$, since there will be no confusion.  Then the quasi-particle energy is given
by
\bea
\omega = \eta(k) \pm \sqrt{\xi^{2}(k)+\Delta^{2}(k)}
\eea
which is identical to that for nodal superconductors except for $\eta(k)$, the imperfect
nesting term.

Here are two urgent questions:

Where can we find UDW?  If they exist, how can we identify them?  As to the second question, we
believe that two hallmarks of UDW are the angle-dependent magnetoresistance and the giant Nernst
effect [\onlinecite{152}].  Both of these are consequences of the Landau quantization of the quasiparticle
spectrum in UDW or Nersesyan's effect.[\onlinecite{153,154}]

\subsection{The Nersesyan effect}

In 1989 Nersesyan et al pointed out that the quasiparticle motion in UDW is quantized
in the presence of a magnetic field perpendicular to the conducting plane.  Then the QP
spectrum becomes
\bea
E_{n}^{\pm}= \eta \pm \sqrt{2nv v_{2}e|B \cos \theta|}
\eea
where n=0,1,2,3... and the $n \neq 0$ states are doubly degenerate and $v_{2}/v=\Delta/E_{F}$ [\onlinecite{155}].
Here $\Delta$ is the maximal energy gap in UDW and we assumed $\eta$ (the chemical potential)
independent of ${\bf k}$.  First the magnetoresistance is given by
\bea
R(B,\theta)^{-1}&=& 2\sigma_{o}+4 \sigma_{1}\left( \frac{e^{-x_{1}}+\cosh(\xi_{0})}{\cosh x_{1}+\cosh \xi_{0}}\right)
\eea
where $x_{1}=\beta \sqrt{2ev v_{2}|B \cos \theta|}, \xi_{0}=\beta\eta$ and $\beta= 1/k_{B}T$.
Here we considered only the 2 lowest Landau levels.  A slight modification of this simple formula
can describe the low-temperature phase of $\alpha$-(ET)$_{2}$KHg(SCN)$_{4}$ [\onlinecite{156}] and the metallic
phase in the Bechgaard salts [\onlinecite{157,158}].  See Fig. 23.
\begin{figure}[ht]
\includegraphics[width=6.75cm]{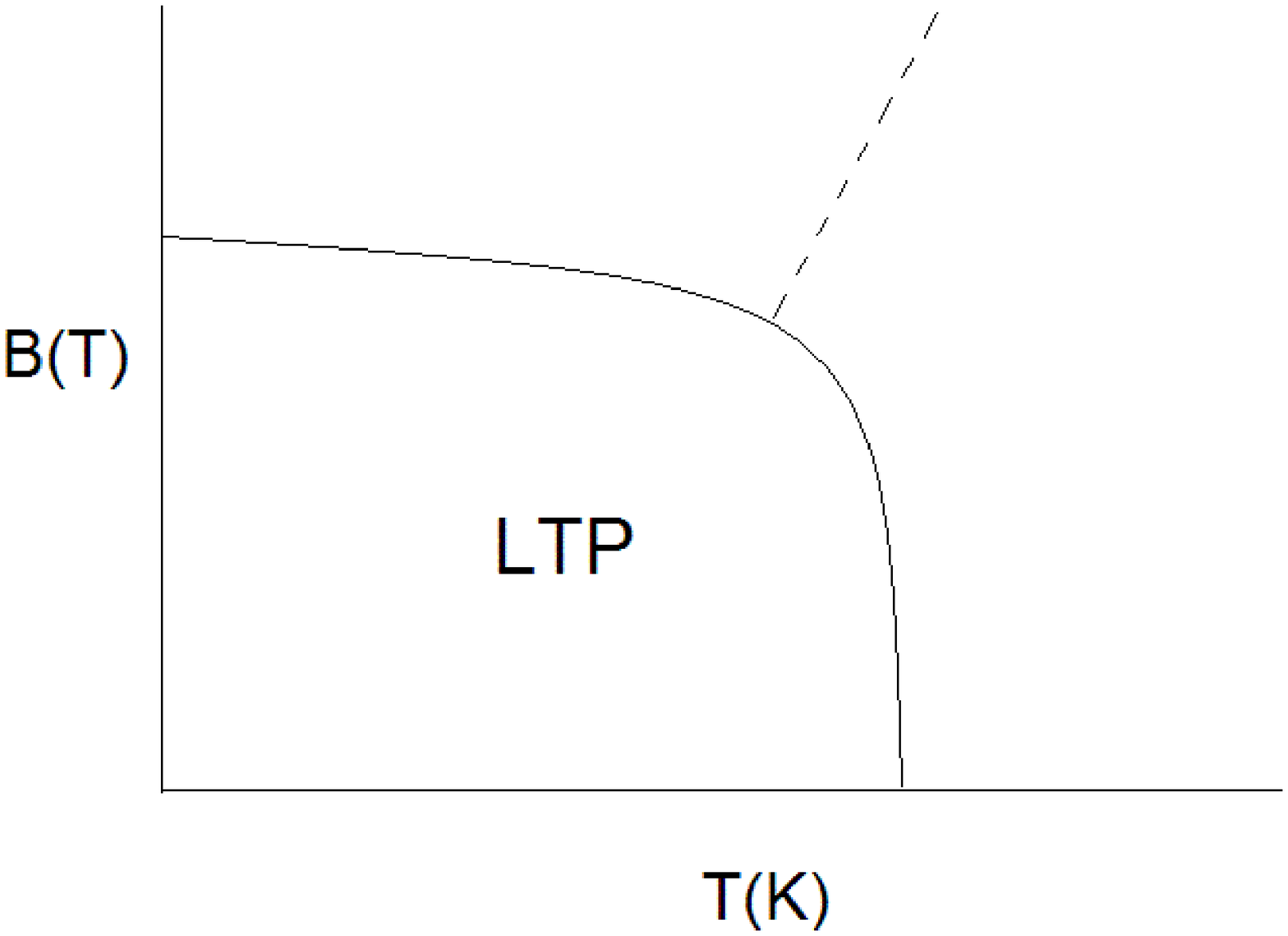}
\includegraphics[width=7.5cm]{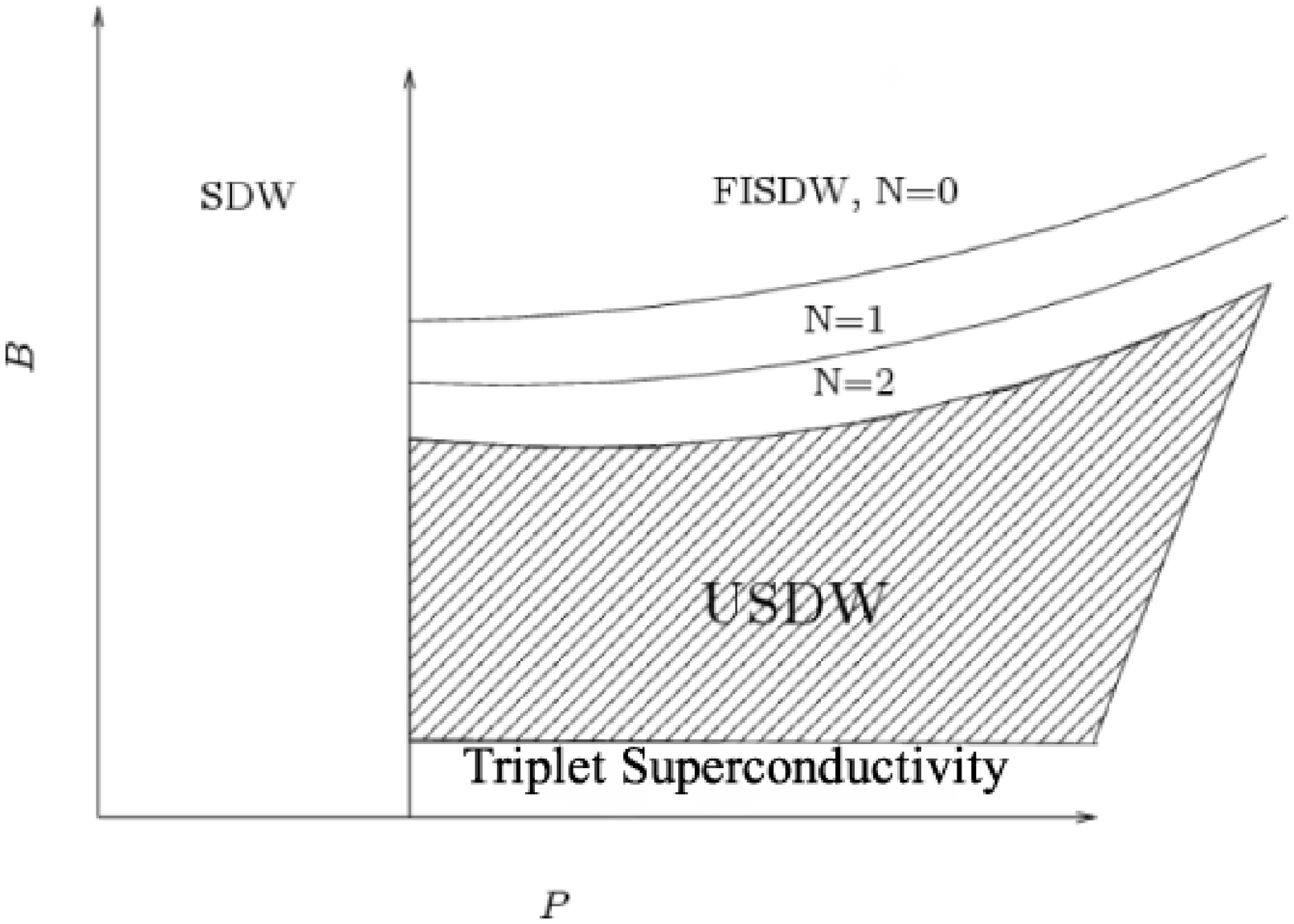}
\caption{Phase diagrams for low-temperature phase of $\alpha$-(ET)$_{2}$KHg(SCN)$_{4}$ 
[\onlinecite{156}](left) and the metallic phase in the Bechgaard salts(right)}
\end{figure}
Second, in the presence of an electric field the QP orbits drift with ${\bf v_{D}}= ({\bf E} {\bf \times} {\bf B})/
|B|^{2}$.  This gives rise to a large negative Nernst effect [\onlinecite{155}].  We obtain
\bea
\alpha_{xy} &=& -\frac{SR}{B} = 2eR(\ln 2 + 2\ln(2\cosh(x_{1}/2))-x_{1}\tanh(x_{1}/2))
\eea
where S is the entropy associated with the QPs.  Indeed a large negative Nernst effect has been 
observed in $\alpha$-(ET)$_{2}$KHg(SCN)$_{4}$ [\onlinecite{159}], in the pseudogap region of high-T$_{c}$ cuprates 
[\onlinecite{160,161,162}],
in the CDW region of NbSe$_{2}$ [\onlinecite{163}] and more recently 
in CeCoIn$_{5}$[\onlinecite{164,165}].  
Recently we have analyzed the Nernst coefficient observed in
CeCoIn$_{5}$ [\onlinecite{165}] in terms of Eq. 105.  This is shown in Fig 24 a) and b).
\begin{figure}
\includegraphics[width=8.5cm]{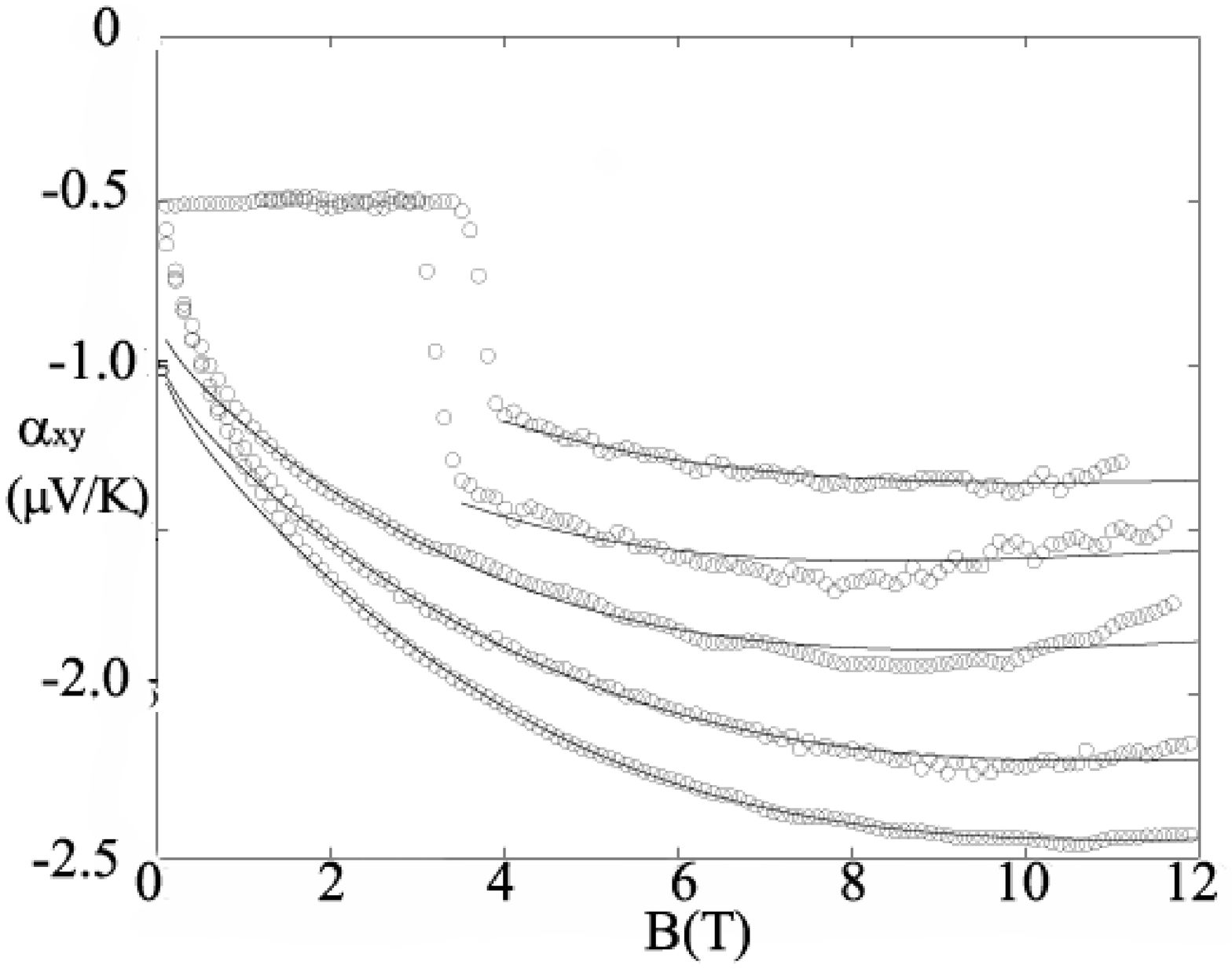}
\includegraphics[width=8.5cm]{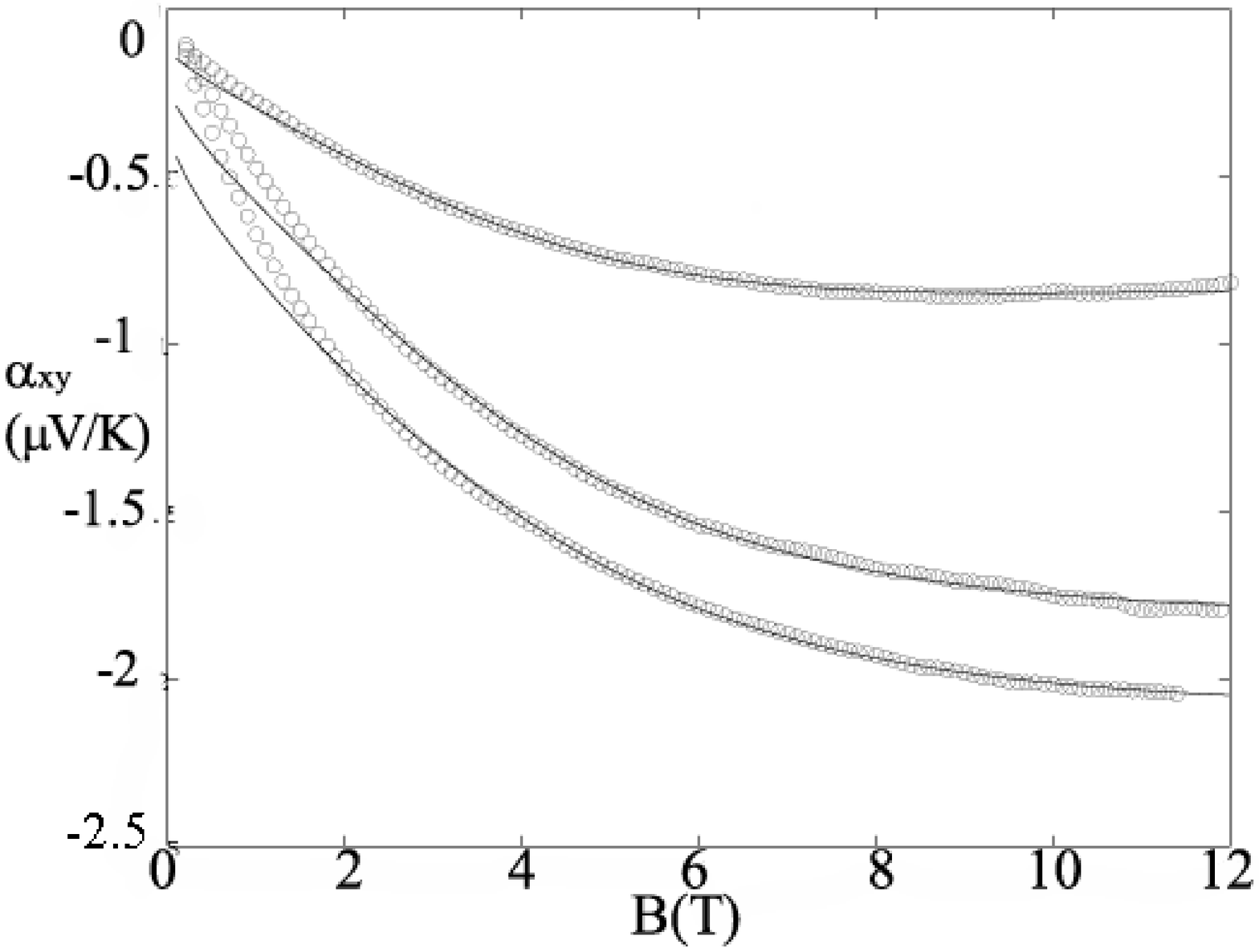}
\caption{The magnetic field dependence of the Nernst coefficient is 
plotted for (left) T = 1.3 K, 1.65 K, 2.5 K, 3.5 K and 4.8 K from top to bottom, and for
(right) T = 7.3 K, 10.5 K and 15 K from bottom to top.  The circles
denote the experimental data, while the solid line is our fit.}
\end{figure}
We obtain an excellent fit as is seen readily.  On the other hand, in order to describe the 
temperature dependence we have to assume
\bea
\Delta(T)v(T) &=& a +bT^{4}
\eea
whose origin is unclear.  However, it is possible that the above temperature dependence
indicates the presence of a quantum critical point (QCP) in CeCoIn$_{5}$.

In addition, we have shown recently the giant
Nernst effect observed in the underdoped region of YBCO, LSCO and Bi-2212 [\onlinecite{160,162}] are described in terms of
dDW [\onlinecite{166}].  Very recently a similar but positive giant Nernst effect has been reported in the AF phase in
URu$_{2}$Si$_{2}$ [\onlinecite{167}].  Indeed there has been a suggestion that CDW in NbSe$_{2}$ is UCDW
[\onlinecite{168}] and the AF phase
in URu$_{2}$Si$_{2}$ is USDW [\onlinecite{169,170}].  So far we have identified 7 candidates for UDW which include the 3 high-T$_{c}$
cuprate superconductors YBCO, LSCO and Bi-2212 in the underdoped region.

Coming back to high-T$_{c}$ cuprates, 1) $\Delta({\bf k}) = \Delta \cos(2\phi)$ in the pseudogap region
is determined by ARPES [\onlinecite{171}].  2) The similar phase diagram as in Fig. 2 is obtained when T is replaced
by the low-temperature energy gap which is measured by STM [\onlinecite{172,173}], ARPES [\onlinecite{damascelli,174}] 
and the universal heat
conduction [\onlinecite{175}].  In particular, if one puts $\Delta= 2.14 T^{*}$ these two phase diagrams match almost 
perfectly.  Needless to say, $\Delta/T_{c} = 2.14$ is also valid for d-wave density wave in the weak-coupling limit
when $|\eta| \ll \Delta$ [\onlinecite{won-1,dora-2}].

\section{Gossamer Superconductivity}

In the first lecture we discussed Laughlin's critique [\onlinecite{laughlin}] on Anderson's RVB state (Eq.(2)).  The 
new wave function suggested by Laughlin can be interpreted as d-wave superconductor in the presence of another
order parameter [\onlinecite{haas,won-2}].  From the phase diagram for high-T$_{c}$ cuprates, the most relevent state is a d-wave
superconductor in the presence of d-wave density wave.  Also this is the prevailing feature in heavy-fermion
systems like URu$_{2}$Si$_{2}$, CeCu$_{2}$Si$_{2}$ [\onlinecite{176}], CeCoIn$_{5}$, CeRhIn$_{5}$, UPd$_{2}$Al$_{3}$,
and organic conductors like $\kappa-(ET)_{2}$ salts [\onlinecite{177,178,pinteric}].

Coming back to high-T$_{c}$ cuprate superconductors, $\Delta({\bf k})$ for d-DW can be readily identified
as the order parameter phenomenologically introduced by Tallon and Loram [\onlinecite{tallon}].  For example, the
superfluid density in the gossamer superconductor at T=0 K is given by
\bea
\rho_{s}(0,\eta) &=& \frac{\Delta_{1}^{2}(0)}{\Delta_{2}^{2}(0)+\Delta_{1}^{2}(0)}
\eea
while the c-axis superfluid density is
\bea
\rho_{s,c}(0,\eta) &=& \frac{\Delta_{1}^{2}(0)}{\sqrt{\Delta_{2}^{2}(0)+\Delta_{1}^{2}(0)}}
\eea
where $\Delta_{1}(0)$ and $\Delta_{2}(0)$ are the maximal gaps of d-wave superconductor and
d-wave density wave, respectively, at T= 0 K.

It is of great interest to explore a variety of transport properties in the gossamer
superconductivity.  For example, if $\eta$, the chemical potential is negligible ($|\eta| \ll T$),
we obtain again the universal heat conduction (Eq.(3)) where now $\Delta=\sqrt{\Delta_{1}^{2}+
\Delta_{2}^{2}}$.  Also, ARPES would see the energy gap [\onlinecite{171}]
\bea
\Delta({\bf k}) &=& \sqrt{(\Delta_{2}\cos(2\phi)-\eta)^{2}+\Delta_{1}^{2}\cos^{2}(2\phi)} \\
& \simeq & |\Delta \cos(2\phi)-\eta|
\eea
If you look carefully at the phase diagram of Bi-2212 (see for example [\onlinecite{174}]), unlike LSCO and
YBCO the superconducting dome in Bi-2212 is completely covered by the pseudogap phase.  This
suggests that the superconductivity in Bi-2212 is gossamer for the whole doping range.  This may
be one of the reasons why $\Delta(0)/T_{c} \sim 5$ in Bi-2212 is so large. In this sense the systematic
study of the optical conductivity, the Raman spectra and the thermal conductivity (with special attention
to the doping dependence of these quantities) is of great interest.

In summary, in exploring unconventional or nodal superconductors, we have encountered a vast forest
inhabited by many unconventional density wave (UDW) and gossamer superconductors where these two
order parameters coexist.  Surprisingly, all of these ground states have been expected from the
infrared instability of the 2D and 3D Fermi liquid.  Therefore we can restore the legacy of
Landau's Fermi liquid theory and BCS theory of superconductivity in a proper perspective.  Then
armed with the Green function methodology as in Abrikosov, Gor'kov and Dzyaloshinski (AGD) [\onlinecite{179}] we
will be fully prepared to explore the plethora of new ground states.

{\bf Acknowledgments}

We have benefitted from close collaboration with Thomas Dahm, Balazs Dora, Koichi Izawa, Konomi
Kamata, Masaru Kato, Hae-Young Kee, Yong Baek Kim, Bojana Korin-Hamzi\'{c}, Yuji Matsuda, 
Peter Thalmeier, Silvia Tomi\'{c} and Attila Virosztek.  Also KM and HW gratefully 
acknowledge the hospitality and the support
of the Max-Planck Institute for the Physics of Complex Systems at Dresden for many summers since 1995.

\end{document}